\documentclass[12pt,letterpaper]{article}

\usepackage{verbatim}
\usepackage{amsmath}
\usepackage{amssymb}
\usepackage{graphicx}
\usepackage{cite}
\usepackage{subfig}
\usepackage{setspace}

\def\Re{{\cal R \mskip-4mu \lower.1ex \hbox{\it e}\,}}
\def\Im{{\cal I \mskip-5mu \lower.1ex \hbox{\it m}\,}}
\def\ie{{\it i.e.}}
\def\eg{{\it e.g.}}

\def\sub#1{_{\lower.25ex\hbox{$\scriptstyle#1$}}}
\def\tev{\,{\rm TeV}}
\def\gev{\,{\rm GeV}}

\def\to{\rightarrow}

\def\subw{_{\rm w}}
\def\mh{\ifmmode m\sbl H \else $m\sbl H$\fi}
\def\mch{\ifmmode m_{H^\pm} \else $m_{H^\pm}$\fi}
\def\mt{\ifmmode m_t\else $m_t$\fi}
\def\mc{\ifmmode m_c\else $m_c$\fi}
\def\mz{\ifmmode M_Z\else $M_Z$\fi}
\def\mw{\ifmmode M_W\else $M_W$\fi}
\def\mws{\ifmmode M_W^2 \else $M_W^2$\fi}
\def\mhs{\ifmmode m_H^2 \else $m_H^2$\fi}   
\def\mzs{\ifmmode M_Z^2 \else $M_Z^2$\fi}
\def\mts{\ifmmode m_t^2 \else $m_t^2$\fi}
\def\mcs{\ifmmode m_c^2 \else $m_c^2$\fi}
\def\mchs{\ifmmode m_{H^\pm}^2 \else $m_{H^\pm}^2$\fi}
\def\ztwo{\ifmmode Z_2\else $Z_2$\fi}
\def\zone{\ifmmode Z_1\else $Z_1$\fi}
\def\mtwo{\ifmmode M_2\else $M_2$\fi}
\def\mone{\ifmmode M_1\else $M_1$\fi}
\def\tb{\ifmmode \tan\beta \else $\tan\beta$\fi}
\def\xw{\ifmmode x\subw\else $x\subw$\fi}
\def\ch{\ifmmode H^\pm \else $H^\pm$\fi}
\def\lum{\ifmmode {\cal L}\else ${\cal L}$\fi}
\def\inpb{\ifmmode {\rm pb}^{-1}\else ${\rm pb}^{-1}$\fi}
\def\infb{\ifmmode {\rm fb}^{-1}\else ${\rm fb}^{-1}$\fi}
\def\epem{\ifmmode e^+e^-\else $e^+e^-$\fi}
\def\ppb{\ifmmode \bar pp\else $\bar pp$\fi}
\def\pbp{\ifmmode ~^(\bar p^)p\else $~^(\bar p^)p$\fi}
\def\bsg{\ifmmode B\to X_s\gamma\else $B\to latexilaX_s\gamma$\fi}
\def\bsll{\ifmmode B\to X_s\ell^+\ell^-\else $B\to X_s\ell^+\ell^-$\fi}
\def\bstt{\ifmmode B\to X_s\tau^+\tau^-\else $B\to X_s\tau^+\tau^-$\fi}

\newskip\zatskip \zatskip=0pt plus0pt minus0pt
\def\matth{\mathsurround=0pt}
\def\lsim{\mathrel{\mathpalette\atversim<}}
\def\gsim{\mathrel{\mathpalette\atversim>}}
\def\atversim#1#2{\lower0.7ex\vbox{\baselineskip\zatskip\lineskip\zatskip
  \lineskiplimit 0pt\ialign{$\matth#1\hfil##\hfil$\crcr#2\crcr\sim\crcr}}}

\renewcommand{\thefootnote}{\fnsymbol{footnote}}

\begin{document} \begin{titlepage} 
\rightline{\vbox{\halign{&#\hfil\cr
&SLAC-PUB-14922\cr
}}}
\vspace{1in} 
\begin{center}

{{\Large\bf The New Look pMSSM with Neutralino and Gravitino LSPs}
\footnote{Work supported by the Department of 
Energy, Contract DE-AC02-76SF00515}\\}
\medskip
\medskip
\normalsize 
{\large Matthew W.~Cahill-Rowley, JoAnne L.~Hewett, Stefan Hoeche, Ahmed Ismail, and Thomas G.~Rizzo\footnote{email: mrowley, hewett, shoeche, aismail, rizzo@slac.stanford.edu} \\
\vskip .6cm
SLAC National Accelerator Laboratory,  \\
2575 Sand Hill Rd, Menlo Park, CA 94025, USA\\}
\vskip .5cm

\end{center} 
\vskip 0.8cm

\begin{abstract} 

The pMSSM provides a broad perspective on SUSY phenomenology. In this paper we 
generate two new, very large, sets of pMSSM models with sparticle masses extending up to 
4 TeV, where the lightest supersymmetric particle (LSP) is either a neutralino or gravitino.
The existence of a gravitino LSP necessitates a detailed study of its cosmological effects and
we find that Big Bang Nucleosynthesis places strong constraints on this scenario. Both sets are
subjected to a global set of theoretical, observational and experimental constraints resulting
in a sample of $\sim 225$k viable models for each LSP type.  The characteristics of these
two model sets are briefly compared.
We confront the neutralino LSP model set with searches for SUSY at 
the 7 TeV LHC using both the missing (MET) and non-missing $E_T$ ATLAS analyses. In the 
MET case, we employ Monte Carlo estimates of the ratios of the SM backgrounds 
at 7 and 8 TeV to rescale the 7 TeV data-driven ATLAS backgrounds to 8 TeV. 
This allows us to determine the pMSSM parameter space coverage for this 
collision energy. We find that an integrated luminosity 
of $\sim 5-20$ fb$^{-1}$ at 8 TeV would yield a substantial increase in this coverage 
compared to that at 7 TeV and can probe roughly half of the model set.  If the pMSSM is not discovered during the 8 TeV run, then our model
set will be essentially void of gluinos and lightest first and second generation squarks that are $\lsim 700-800$ GeV, 
which is much less than the analogous mSUGRA bound. Finally, we demonstrate 
that non-MET SUSY searches continue to play an important role in exploring the 
pMSSM parameter space. These two pMSSM model sets can be used as the basis for investigations for years to come.

\end{abstract}

\renewcommand{\thefootnote}{\arabic{footnote}} \end{titlepage}


\section{Introduction and Background}
\label{sec:intro}

Supersymmetry (SUSY) is a favored paradigm {\cite{SUSY}} for physics beyond the Standard Model (SM). The reasons for this are well-known: 
($i$) SUSY predicts the unification of the gauge couplings, ($ii$) SUSY stabilizes the hierarchy, ($iii$) SUSY has several attractive dark matter  
candidates, and, perhaps most importantly, ($iv$) SUSY can be directly searched for and tested in collider (as well as other) experiments. Since SUSY 
stabilizes the electroweak scale at $\sim 246$ GeV one expects the SUSY partners of the SM fields to have masses not too far from this value. 
Unfortunately, while searches for the Higgs boson at the LHC by both the ATLAS{\cite{ATLAS:2012ae}} and CMS{\cite{Chatrchyan:2012tx}} Collaborations have yielded 
tantalizing hints for a signal in the $\simeq 123-127$ GeV mass region, the analogous searches for supersymmetric particles {\cite {lathuile}} 
have so far been unsuccessful and hint that most of the colored part of the SUSY spectrum may be relatively heavy.  It is thus imperative to explore
the consequences of these search results in various SUSY scenarios.

There are several approaches for examining the predictions of SUSY within the Minimal Supersymmetric Standard Model (MSSM) framework. 
Perhaps the most popular is to assume a specific UV SUSY-breaking scenario, such as minimal Supergravity/constrained MSSM (mSUGRA/CMSSM){\cite {msugrab}} 
or gauge-mediated SUSY breaking (GMSB) {\cite {gmsbb}}, which have the advantage of having only a few parameters. However, such scenarios can suffer 
from restricted flexibility when constraints from different experiments are combined, leading to small viable parameter space regions with limited signatures. 
A second possibility is to employ Simplified Models{\cite {simplified}}, which allow for each SUSY search to directly constrain the relevant sparticle masses, but do not allow for understanding correlations between data from different (\eg, collider and dark matter) sectors. A third approach is to consider the p(henomenological) MSSM{\cite{Djouadi:2002ze}}, which we investigate here, following
our earlier work{\cite {us}}. Depending on the specific version employed, the pMSSM is described by $\sim 20$ real 
soft-breaking parameters that are defined at the TeV scale. This parameter set is large enough to provide a high degree of flexibility, yet small enough that correlations between different types of searches can be examined with a (rather large but) finite amount of computing resources. Although these resources are currently 
insufficient to fully explore this very large parameter space, the analyses that have been performed so far reveal the 
many-fold possibilities of the full MSSM. 

The impact of ATLAS and CMS SUSY searches at the 7 TeV LHC with $\sim 1$ fb$^{-1}$ or more of luminosity on the pMSSM parameter space has now been considered by 
several authors{\cite {us,them}}. The result of these studies is that, except for a few potential holes near kinematic boundaries, gluinos and the first/second 
generation of squarks must be fairly massive. A large fraction of these previous studies, including our own{\cite {us}}, focused on the possibility of 
relatively light sparticles with all masses in the SUSY spectrum being below $\sim 1$ TeV which now seems to be disfavored (though not yet excluded). Our earlier results {\cite {us}}, as well as the parallel studies 
of these same model sets{\cite{Strubig:2012qd}}, both conclude that nearly all ($\sim 99\%$) of such models are now excluded by the 
ATLAS and CMS SUSY searches. This being the case, it is well motivated to generate new sets of pMSSM models with soft SUSY breaking parameters above 
1 TeV, that can not be so easily excluded in the near future, and can be used for LHC and other studies.

In this paper we will extend our previous analysis and discussion of pMSSM models{\cite {us}} in several ways. First, we will now consider the possibility 
that either the lightest neutralino or the gravitino is the lightest supersymmetric particle (LSP) and contributes to the observed dark matter relic density, and we
will generate a set of viable models for each of these LSP scenarios. 
As will be discussed in detail below, this  
choice will have a significant impact on the procedure used to generate these two separate model sets. In particular, we pay careful attention to the restrictions that
arise from cosmological considerations, such as Big Bang Nucleosynthesis, in the case of a gravitino LSP.
Second, as per the above discussion, we will  
allow all of the soft mass parameters to take on values as large as 4 TeV. This larger range will allow for SUSY studies to be performed deep into the future 13-14 TeV run 
of the LHC. Third, since the results of ATLAS SUSY missing transverse energy (MET) analyses, which we closely follow, are now available for an integrated 
luminosity of $\sim 1$ fb$^{-1}$ and, in some cases, for $\sim 5$ fb$^{-1}$, we can make use of their data and benchmark models to accurately validate our own analyses and results. Note that in our earlier work, we 
could only make use of ATLAS pre-data Monte Carlo (MC) studies{\cite {atlasmc}}.  Here, we investigate the ATLAS search reach for our new neutralino LSP pMSSM
model set from the 7 TeV run, and also explore the sensitivity at 8 TeV.   As will be discussed at length below, the 7 TeV data can be 
used to make reasonable extrapolations to results that may be obtained during the present 8 TeV run. Amongst other things, such an extrapolation requires an estimate of the SM 
backgrounds at 8 TeV based on the ATLAS data-driven backgrounds at 7 TeV. In our analysis below, we employ a MC approach to calculate the anticipated 
{\it ratios} of the 7 and 8 TeV backgrounds in each of the relevant channels and then rescale the appropriate ATLAS-obtained 7 TeV backgrounds correspondingly. Although 
MC estimates of SM backgrounds at a fixed $\sqrt s$ are at best approximate (hence the use of data-driven methods), we expect a SM MC 
analysis to perform well in obtaining background ratios at energies that are not too far apart. Applying these techniques to the new neutralino 
LSP model set, we will show that the 8 TeV ATLAS SUSY searches with $\sim 20$ fb$^{-1}$ of luminosity will cover somewhat more of the pMSSM parameter space 
than do the present 7 TeV searches. However, a truly significant step to reach a more complete coverage will require going to the full LHC energy of 13-14 TeV.  The implications of LHC searches for the 
gravitino model set will be discussed in a future work{\cite {future}}, as the analysis is somewhat different from the neutralino LSP case. 
 
The outline of this paper is as follows. In Sections~\ref{sec:modgen} and~\ref{sec:cosmo},  we describe our procedure for generating both the neutralino and gravitino LSP 
pMSSM model sets. We pay special attention to the cosmological and astrophysical considerations that we have employed in these two cases as it is here that 
the constraints most differ.  In Section~\ref{sec:compare} we compare the properties of the two model sets before applying the results from 
the ATLAS MET analyses. The ATLAS MET analyses and their extrapolations to 8 TeV for the neutralino LSP 
model set will be discussed in Section~\ref{sec:met}, while some non-MET based searches will be discussed in Section~\ref{sec:nonmet}. Further discussion and our conclusions can be found in Section~\ref{sec:conc}.

\section{Model Generation}
\label{sec:modgen}

Here, we describe the procedure we follow in generating our large sample of models, consistent with the global data set, for the pMSSM with both a neutralino and gravitino LSP.

\subsection{Neutralino LSP}
\label{sec:modgen-neut}

We study the 19-dimensional parameter space of the pMSSM \cite{us}.  This parameter set results from imposing the following minimal assumptions on the general MSSM:
($i$) R-parity is taken to be conserved, ($ii$) the soft parameters are taken to be real so that no new sources of CP violation exist beyond that present in the CKM matrix,
($iii$) Minimal Flavor Violation~\cite{mfv} is taken to be valid at the electroweak scale, ($iv$) the first 2 generations of squarks and sleptons with the same quantum numbers
are taken to be degenerate and to have negligible Yukawa couplings.  In this case, we also assume that the lightest supersymmetric particle is the lightest neutralino and
that it undergoes a thermal cosmology consistent with being a stable dark matter candidate.  Effects of physics at high energy scales, or associated with a specific SUSY breaking source,
are not considered.  In particular, we emphasize that we do not assume a fixed SUSY breaking mechanism.  The last three of these conditions ensure compliance with restrictions in the flavor
sector.  These assumptions reduce the large MSSM parameter space to the 19 parameters considered here.

In generating our model set, we first perform a numerical scan over these parameters, which are defined at the TeV scale.  $3\times 10^6$ points in the 19-dimensional parameter space were randomly 
generated assuming flat priors, {\it i.e.}, the parameter values were uniformly chosen over their scan range.  We found in our earlier work that the choice of logarithmic priors
did not significantly modify the results \cite{us}. The 19 parameters, and our scan ranges, are given by
\begin{eqnarray}
100 \gev & \le & m_{\tilde L(e)_{1,2,3}} \le 4 \tev \,\nonumber\\
400 \gev & \le & m_{\tilde Q(q)_{1,2}} \le 4 \tev \,\nonumber\\
200 \gev & \le & m_{\tilde Q(q)_{3}} \le 4 \tev \,\nonumber\\
50 \gev & \le & |M_1| \le 4 \tev \,\nonumber\\
100 \gev & \le & |M_2, \mu| \le 4 \tev \, \label{eqn:params}\\
400 \gev & \le & M_3 \le 4 \tev \,\nonumber\\
|A_{t,b,\tau}| & \le & 4 \tev \,\nonumber\\
100 \gev & \le & M_A \le 4 \tev \,\nonumber\\
1 & \le & \tan \beta \le 60 \,\nonumber
\end{eqnarray}
The absolute value signs are present to allow the soft-breaking parameters to have arbitrary sign.
The lower bound on these ranges were chosen to maximize consistency with LEP II, Tevatron, and LHC7 SUSY searches~\cite{Nakamura:2010zzi}.  Note that this implies a lower bound of $\sim 400$ GeV for the gluino
and the first two generation squark masses, and safely sets the lightest stop to be heavier than the top-quark.  The upper end of the scan range was set by the approximate kinematical
reach of the LHC operating at 14 TeV.  In this case, we expect some fraction of our models to remain viable after the 7-8 LHC runs and thus our model set will provide a useful base
for studies in the years to come.

For each set of weak scale parameters, the resulting SUSY spectrum was generated via SOFTSUSY 3.1.7~\cite{Allanach:2001kg}.  We performed a cross-check of this spectrum calculation via the numerical
package Suspect~\cite{Djouadi:2002ze}.  Any models where discrepancies for the sparticle masses between the two generators occurred at the order of 25\% or more were discarded; we, surprisingly, found a few such cases which totaled a small fraction of our model sample.  The sparticle decays were computed using a modified version of SUSY-HIT 1.3~\cite{Djouadi:2006bz}.
Our modifications include ($i$) the incorporation of the light quark and lepton masses in the calculation of
branching fractions and lifetimes for the various sparticles. Two-body decays implement the full mass corrections, while 3-body decays only include a modified phase-space cut-off.  
We also included the mass of the lightest meson of the appropriate type in the relevant phase space calculations to account for hadronization effects.  ($ii$) One-loop and CKM
suppressed decays were included for the case of sbottom squarks.  ($iii$) Full expressions were employed for chargino decays which are important when the chargino-neutralino mass
splitting is $\lsim 1$ GeV.  ($iv$) Four-body final states were included in the decays of stops with small mass splittings, incorporating the appropriate final state fermion masses.
($v$) The decay width of sparticles with boosted decay lengths greater than $\sim 4$ m was set to zero to avoid issues with collider event generation.  ($vi$) We removed the QCD corrections to decays involving stops and sbottoms due to their tendency to result in negative decay widths.  ($vii$) We added the tau sneutrino mass to HDECAY, since the public version assumes all sneutrinos to be degenerate. Modifications $i$ - $v$ are described in detail in our previous work \cite{us}.

Once a model ({\i.e.}, a set of parameters) has been generated, we subject it to a global set of theoretical and experimental constraints to ensure its validity.  We briefly
enumerate these restrictions here and further details can be found in \cite{us}.  ($i$) On the theoretical side,
we demand that the spectra be tachyon free, the models contain no color or charge breaking minima, and that a bounded Higgs potential must exist.  ($ii$) We employed a number of
constraints from the flavor sector and the precision electroweak data arising from the measurements of $(g-2)_\mu$, $b\to s\gamma$, $B_s\to\mu^+\mu^-$, meson-anti-meson mixing,
the invisible width of the $Z$-boson, and the isospin splitting parameter $\Delta\rho$.  The rare decay $B_s\to\mu^+\mu^-$ is of particular interest and will be discussed
further below.  ($iii$)  Sparticle and Higgs searches from LEP were imposed.  Some of these searches have been re-evaluated to remove model dependent assumptions.  ($iv$) Lastly,
we have required that the neutralino LSP contribution to the dark matter relic density not exceed the upper bound determined by WMAP~\cite{Komatsu:2010fb}, and that the LSP
scattering cross sections be consistent with the restrictions from dark matter direct detection.  Note that we have chosen a lower limit for our scan range of the weak scale
parameter $M_1$ of 50 GeV in order to avoid the current confusing situation regarding direct detection results for low WIMP masses~\cite{Gelmini:2011xz}.
 
After imposing this set of theoretical and experimental constraints, 223,256 viable models remain to form our set.  We are now set to study the characteristics
of this model sample and their signatures at the LHC.

\subsection{Gravitino LSP}
\label{sec:modgen-grav}


Our scan of the gravitino pMSSM is based on the framework discussed above that was developed in~\cite{us} for the pMSSM with a neutralino LSP.  Due to their weak interactions, gravitinos decouple from most of our calculations, and we can often use the same numerical procedure in both the neutralino and gravitino scans. Differences between the two scans arise chiefly from three sources: First, the lightest neutralino is no longer required to be lighter than the other sparticles, allowing almost any SUSY spectrum to be viable which in turn results in a large variety of possible decay pathways. Second, the gravitino becomes a dark matter candidate. Gravitino dark matter differs from the neutralino case both in its relic abundance and in its inaccessibility to direct detection searches. Finally, the NLSP (and potentially other sparticles) will decay directly to the gravitino. This leads to potentially small decay widths of the NLSP and other sparticles which result in collider and cosmological signatures that are quite distinct from those found in the neutralino pMSSM.  


The first step of our analysis is to add the mass of the gravitino, the spin-3/2 superpartner of the graviton, to our input parameters. In the absence of supersymmetry breaking, the gravitino would be massless and would have only gravitational-strength interactions. However, supersymmetry breaking gives the gravitino a mass and enhances its interactions with the MSSM. In our study, we consider models in which F-term supersymmetry breaking results in a massless goldstino with interactions determined by the supersymmetry-breaking vev, $\langle F \rangle$. When the gravitino eats the goldstino, it acquires a mass $M_{\tilde{G}} = \frac{\langle F \rangle}{\sqrt{3} M_P}$. Since gravitinos couple to the rest of the MSSM with a dimensionful coupling of $\langle F\rangle^{-1/2}$, the interactions and therefore phenomenology of a gravitino is determined entirely by its mass. For our parameter scan, we add the scale of supersymmetry breaking, $\sqrt{\langle F \rangle}$, to the set of pMSSM parameters described in Eqn.~\ref{eqn:params}. To cover a variety of gravitino masses, we logarithmically sample the range $6.45 \times 10^4$ GeV $< \sqrt{\langle F \rangle} < 6.45 \times 10^{10}$ GeV, corresponding to the mass range 1 eV $< M_{\tilde{G}} < 1$ TeV.

The lower limit for the scan range for the gravitino mass results from several considerations. First, models where the gravitino mass is less than 1 eV have prompt decays, so smaller gravitino masses have a minimal impact on the collider phenomenology, and our model set therefore captures most of the interesting physics resulting from a light gravitino LSP. We note that models with extremely light gravitinos can be detected by mono-jet searches at the LHC~\cite{Brignole:1998me}, which we do not consider here. Second, gravitinos with masses below 1 eV are cosmologically uninteresting due to their insignificant relic density. Finally, a gravitino mass of 1 eV provides a comfortable separation between the TeV scale and the scale of supersymmetry breaking, so that we do not have to worry about the effects of the physics responsible for supersymmetry breaking. The upper bound on the gravitino mass scan range is chosen to cover the complete range where the gravitino can be the LSP. While a gravitino in our model set could theoretically have any mass below 4 TeV, the heaviest gravitino in our model set is 760 GeV, indicating that the 1 TeV upper limit is not saturated. 


After choosing the weak-scale parameters for a model, we generate the physical spectrum which is unaltered by the addition of a gravitino.  Since the gravitino interacts far too weakly to affect the other sparticle masses, any theoretically-consistent spectrum may in general be allowed. This contrasts sharply with the neutralino scan, in which the LSP mass is related to the properties of the other neutralinos and charginos, which are important for the phenomenology of the model. Removing the requirement that the LSP be the lightest neutralino results in a significantly different distribution of NLSP types, as we will see below. 

Once a model spectrum is generated, we check for consistency with the long list of global experimental and theoretical constraints described above. Weak couplings prevent gravitinos in our mass range from altering theoretical constraints on the vacuum, values of electroweak precision observables, rare decay rates, and LEP Higgs searches. However, the presence of a gravitino LSP has a strong impact on stable charged particle searches and precision cosmology, which place additional restrictions on the
model parameters.


Limits on stable charged particles apply identically to both the neutralino and gravitino model sets. However, they are considerably more important for gravitino models, in which the NLSP can only decay to a gravitino. Depending on the gravitino mass, the NLSP may have a long lifetime, leading to displaced vertex or stable particle signatures. Sparticles that are not the NLSP may also be seen in displaced vertex or stable particle searches if their decays to lighter sparticles are suppressed by a combination of phase space and a high final state multiplicity. While a full treatment of long-lived particle searches will be included in a future study of the LHC phenomenology of our gravitino model set~\cite{future}, we apply here the stable particle limits developed for the neutralino scan as a precursor to a more detailed analysis. This analysis is complicated by the variety of potentially dominant decay channels for a given sparticle, including the canonical decay modes present in the neutralino model set, direct decays to the gravitino, and non-canonical decays to another sparticle (often the NLSP).  The computation of sparticle decays thus requires careful treatment, which we now describe.

Two-body decays to gravitinos are calculated using a modified form of SUSY-HIT based on version 1.2~\cite{Djouadi:2006bz}. The public code calculates decays of neutralino, chargino, gluino, and stau NLSPs, via the approximate formula $\Gamma_{\tilde{X} \to x \tilde{G}}  = K \times \frac{M_{\tilde{X}}^5}{48 \pi M_{Pl}^2 M_{\tilde{G}}^2}$. ($K$ is an ${\cal O}(1)$ function of mixing angles that depends on the identity of $\tilde{X}$). For our study, we modified this to calculate the gravitino decay width for all sparticles and used exact analytic formulae where neither the gravitino nor the Standard Model particle are assumed to be massless. Formulae for neutralino, chargino, and gluino decays can be found in~\cite{Feng:2004mt}, while sfermion decays are calculated using the decay width for a stop NLSP calculated in~\cite{Santoso:2007uw}. When the NLSP is a stop, chargino, or neutralino, the two-body decay may be kinematically forbidden, requiring the calculation of a three-body decay where the Standard Model particle is off shell. We calculate these decays with MadGraph 4~\cite{Alwall:2007st}, modified as described in~\cite{Hagiwara:2010pi} to include the full spin-3/2 sum rules that allow for accurate width calculations for gravitinos of any mass. This is important because the standard goldstino approximation, in which only the spin-1/2 modes of the gravitino are considered, is not valid when the gravitino mass is similar to the mass of the decaying sparticle.

In the gravitino model set, sparticles typically undergo cascade decays ending with the NLSP, which then decays to its Standard Model partner and a gravitino.  In some cases, however, decays to four and five-body final states may dominate.  In these cases, we use CalcHEP~\cite{calchep} to numerically evaluate the available three, four, and five-body decay widths. Examples where such decays can be important are $\tilde{e}_R \to \tilde{u}_R \: \bar{u} \: e$, $\tilde{g} \to \tilde{e}_L \: \bar{e} \: u \: \bar{u}$, and $\tilde{t}_1 \to \tilde{e}_R \: \bar{e} \: u \: \bar{d} \: b$, where the last channel is only relevant when the splitting between the stop and right-handed selectron is less than $M_W + M_b$. Combining this with our modified version of SUSY-HIT accurately computes the largest tree-level decay width for any sparticle heaver than $M_b + M_{NLSP}$. We include the effects for splittings below the $b$ mass in the decays of charginos where the chargino is very close in mass to a neutralino NLSP since this particular scenario is quite common in our model set. 

Finally, we note that the common occurrence of high-multiplicity final states suggests that radiative decay modes may be important. Implementing the large number of potential radiative decay modes, beyond those currently included in SUSY-HIT, is beyond the scope of this work. We expect that the rate for three-body radiative decays would similar in magnitude to that for the five-body tree-level decays, and hence we believe we are not omitting any sizable decay channels. 

\section{Gravitino Cosmology}
\label{sec:cosmo}


Here, we consider the cosmological significance of gravitino LSPs in the 1 eV to 1 TeV mass range, and discuss how we apply cosmological constraints to the gravitino LSP model set. Supersymmetric models with a gravitino LSP are constrained by a variety of cosmological observations and these considerations differ substantially from the case of a neutralino LSP. Since we assume R-parity conservation, gravitino LSPs are stable and must have a relic density below the upper limit from WMAP 7, $\Omega h^2 \lsim 0.1234$~\cite{Komatsu:2010fb}. Additionally, when the gravitino is heavier than about 1 MeV, the NLSP can decay late enough ($\tau_{NLSP} \gtrsim 10^{-2}$ s) to affect the abundances of light elements produced during Big Bang Nucleosynthesis (BBN). The decay products of extremely long-lived NLSPs ($\tau_{NLSP} \gtrsim 10^{7}$ s) can alter spectral parameters of the Cosmic Microwave Background (CMB), affect the reionization history of the universe, or be directly detected as a diffuse background of photons or neutrinos. Cosmological observations can therefore be used to place an upper bound on the energy density of the NLSP before it decays. In general, this bound depends on the NLSP lifetime, interactions, and decay products. The properties of the NLSP and gravitino may also have observable consequences for structure formation. 

These cosmological constraints (other than the WMAP bound on the relic density) can be parameterized as limits on the NLSP density as a function of its lifetime, and depend strongly on the decay modes of the NLSP. The properties of the decay products can be described by the electromagnetic and hadronic branching fractions, the mass of the NLSP, and the fraction of energy carried away by the gravitino. 

The electromagnetic branching fraction is 100\% for all NLSPs except sneutrinos. (We ignore the small invisible width of neutralinos resulting from invisible decays of the Z.) For sneutrinos, both products of the dominant 2-body decay are invisible, and hence only 3-body and 4-body decays can produce electromagnetically-interacting particles. Decays to light gravitinos ($M_{\tilde{G}} \lesssim 1$ GeV) may be calculated in the goldstino limit using CalcHEP~\cite{calchep}. However, decays to heavier gravitinos require use of the full spin-3/2 sum rules and are computed using the spin-3/2 extension of MadGraph~\cite{Hagiwara:2010pi} described previously. In some cases, weak-scale bosons are kinematically forbidden and a four-body decay through a virtual boson is required to produce electromagnetically-interacting particles. In this case, the electromagnetic branching fraction may be highly suppressed, allowing sneutrinos to inhabit regions of parameter space unavailable to other meta-stable NLSPs.

The hadronic branching fraction $B_h$ is of paramount importance for NLSPs decaying between 10$^{-2}$ seconds and 10$^5$ seconds. In most cases we can simply compute the fraction of NLSP decays that result in colored particle production. Here, we calculate 2- or 3-body decay widths to real weak bosons, then multiply by the hadronic branching fraction of the weak boson. In some cases, the weak bosons are virtual, and we calculate the full 3- or 4-body decay widths, using the spin-3/2 sum rules 
in MadGraph to ensure the accuracy of the decay width when the goldstino approximation fails. 

However, in some cases the hadronic decay products contain a much higher fraction of mesons than the 100 GeV shower used to calculate the BBN constraints we apply to our models. We conservatively assume that the extra mesons do not affect BBN, which is accurate for NLSPs with a lifetime $\gsim 10$ seconds since mesons produced after this time decay before interacting~\cite{Jedamzik:2006xz}. The practical effect of this assumption is to count tau leptons only as electromagnetic energy and to require calculation of an effective hadronic branching fraction for neutralinos decaying through a virtual photon, which tend to produce very soft $q \bar{q}$ pairs~\cite{Kawasaki:2004qu}. This effective hadronic branching fraction removes the mesonic decay products by weighting decay modes according to the average number of baryons they produce, so that soft $q \bar{q}$ pairs are not counted as hadronic energy.

When applying cosmological constraints to our pMSSM model set, we include an additional factor of 2 to account for uncertainties in the constraints, our calculated branching fractions, and the NLSP density. We thus feel that our resulting bounds are quite conservative, in the sense that we exclude models cautiously. In the following, we describe how cosmological constraints are implemented and discuss their impact on our model set.

\subsection{Relic Density of Gravitinos}

The relic density of gravitinos is determined by both thermal and non-thermal processes. Thermal production occurs through the standard freeze-out mechanism, and tends to overclose the universe for gravitinos heavier than $\sim$ 2 keV~\cite{Moroi:1993mb}. One way to avoid excess gravitino production is by assuming a low reheating temperature, so that gravitinos do not achieve thermal equilibrium after inflation. Requiring that the gravitino relic density is less than the observed value results in a limit on the reheating temperature, which ranges from $10^{10}$ GeV for a 10 GeV gravitino to an uncomfortably low 100 GeV for a 2 keV gravitino~\cite{Moroi:1993mb}. However, plausible mechanisms have been proposed to dilute the gravitino relic density through entropy addition, thus relaxing or eliminating the upper limit on the reheating temperature (see, for example,~\cite{Fujii:2002fv}). In order to be as general as possible, we therefore place no constraints on the thermal relic density of gravitinos. 

Non-thermal production of gravitinos results from out-of-equilibrium NLSP decays. When the gravitino is heavier than $\sim$ 100 keV, NLSP decays are too slow to affect the thermal production of the NLSP, which therefore freezes out before decaying~\cite{Moroi:1993mb}. The resulting relic density of gravitinos can be computed by assuming that the NLSP is stable, calculating its relic density, and then rescaling by the energy lost during NLSP decay.  The relic density of a meta-stable NLSP can essentially be computed in the same manner as for a neutralino LSP, by numerically evolving the densities of the NLSP and other sparticles using thermally averaged production and annihilation cross-sections. We perform this evolution using micrOMEGAs 2.4~\cite{micromegas}. Since the gravitino is decoupled from the thermal bath for all temperatures of interest, we neglect it and treat the NLSP as a cosmologically stable particle regardless of its electric or color charge. 

While electrically charged particles freeze out in the same manner as neutral ones, colored NLSPs present an additional challenge. The complication arises from interactions of colored NLSPs below the confinement temperature, which are described by poorly-understood non-perturbative strong dynamics \cite{qcdrelic, Kusakabe:2011hk}.  Broadly speaking, colored NLSPs hadronize with quarks and gluons, forming heavy hadrons which then combine to form bound states containing multiple NLSPs. NLSPs in these bound states may co-annihilate, or annihilate with NLSPs from another bound state, severely depleting their abundance. Alternatively, the NLSPs within a bound state may be unable to co-annihilate and then interact weakly with other bound states, maintaining a density similar to the perturbatively-calculated value.  Estimates~\cite{Kusakabe:2011hk} of the relic density of a stable colored particle span a wide range of values, $4 \times 10^{-7} < \Omega h^2 < 6 \times 10^{-3} $ for particle masses between 500 GeV and 3 TeV, depending on the species formed by hadronization.
Although the effects of hadronization on the relic density of colored NLSPs are only understood qualitatively and may be strongly model-dependent, our model set is nearly unaffected by this uncertainty. The reason for this insensitivity is that BBN excludes colored particles with lifetimes longer than $\sim$ 200 s even when the NLSP density is extremely suppressed, as we will see below.

Once we have calculated the non-thermal gravitino relic density, we require it to be less than the WMAP upper limit, $\Omega_{\tilde{G}} h^2 = \Omega_{NLSP} h^2 \times \frac{M_{\tilde{G}}}{M_{NLSP}} < 0.1234$. 

\subsection{Big Bang Nucleosynthesis}

By far the strongest cosmological constraints on our gravitino model set arise from BBN. Energetic particles from NLSP decays can alter the non-equilibrium abundances of nucleons and nuclei, modifying the resulting abundances of light elements~\cite{BBN}. Astrophysical measurements of primordial abundances are generally consistent with the standard model of BBN, and can be used to set limits on any alterations to the standard scenario. The mechanism that alters elemental abundances depends on the time at which the NLSP decays, which in turn determines the fate of its decay products and the nuclear species present at the time of energy injection. Only particles with lifetimes longer than 10$^{-2}$ s can affect BBN since nucleon abundances at earlier times are determined by thermal equilibrium~\cite{Jedamzik:2006xz}. Before $\sim 10^5$ s, only hadronic decay products are important since $\gamma$-rays and electrons loose their energy through interactions with the CMB before scattering on nucleons. After this time, all electromagnetically-interacting decay products have an equal effect, which depends only on their total energy. 

In~\cite{Jedamzik:2006xz}, conservative ranges for the primordial abundances of $D$, $^3He$, $^4He$, $^6Li$, and $^7Li$ are used to derive constraints on an unstable neutral particle, $X$, with a lifetime between $10^{-2}$ s and $10^{12}$ s. Constraints are given in terms of the relic density the decaying particle would have if it was stable, 
$\Omega^*_{NLSP}h^2$, as a function of its lifetime. These constraints are derived for a wide range of hadronic branching fractions and for two masses, 100 GeV and 1 TeV. Hadronic decays are assumed to be of the form $X \to q \bar{q}$, and are showered in PYTHIA to produce the final decay products. Since the NLSP decay chains in our pMSSM model set differ from this generic scenario, we must adapt the tabulated constraints before applying them to our model set as described below. 

For neutral NLSPs, we adapt the exclusion contours provided in~\cite{Jedamzik:2006xz} for a 100 GeV particle decaying to $q \bar{q}$. We rescale these contours by the electromagnetic branching fraction ($\approx 1$ for all NLSPs except sneutrinos, for which it can be extremely small). Since hadronic and electromagnetic energy injection affect BBN differently, we distinguish between hadronic and electromagnetic constraints when applying them to our models. 

In our model set, hadronic decays generally occur indirectly through the production of an on-shell weak boson, so NLSP decays generally produce the same number of baryons as a 100 GeV shower, but with an energy closer to 1 TeV. As a result, we must make some assumption in deciding how to apply the constraints from~\cite{Jedamzik:2006xz}. We choose to assume that the effect on BBN depends entirely on the number of baryons produced, which is valid if the produced baryons thermalize before interacting hadronically. Since an NLSP decaying through a weak boson produces a fixed number of baryons, the constraint on the NLSP {\it number} density would be independent of the NLSP mass in this case. The constraint on the {\it energy} density would therefore become weaker with increasing mass. To apply this assumption, we simply use the 100 GeV limits to determine the maximum number density of injected weak bosons, which is also the maximum number density of the NLSP. We choose this assumption for two reasons. First, it is the most conservative possibility - it ignores the extra energy of a heavy NLSP, which would otherwise lead to stronger constraints. Second, it is supported by considering the difference in the tabulated 100 GeV and 1 TeV limits. At early times, the 1 TeV limit is weaker by a factor of $\sim$ 5, which we can explain from the fact that the generic 1 TeV particle (which decays directly to $q \bar{q}$) produces approximately twice as many baryons per decay as the 100 GeV particle (estimated from Figure 9 of \cite{Kawasaki:2004qu}). The number density constraint on the 1 TeV particle should therefore be twice as strong, leading to an energy density constraint that is weaker by a factor of five. At times approaching the transition to electromagnetic constraints, the 100 GeV and 1 TeV constraints move closer together, indicating that the extra energy of the 1 TeV particle is becoming relevant and therefore that our approximation is becoming increasingly conservative.

In the region where only the electromagnetic branching fraction is important, we need not worry about the decay product composition, since the constraints depend only on the total visible energy injected. However, we must calculate how much energy is imparted to electromagnetically-interacting decay products. We conservatively assume that the gravitino carries away a fraction $\frac{1}{2}(1+(\frac{M_{\tilde{G}}}{M_{NLSP}})^2)$ of the energy (more energy remains, leading to stronger constraints, if the Standard Model decay products have a significant mass). Since the NLSP lifetime increases as the gravitino becomes heavy, the effect of this energy loss is to weaken the BBN constraints on extremely long-lived NLSPs. 

\begin{figure}
\centering
\subfloat{\includegraphics[width=6.5in]{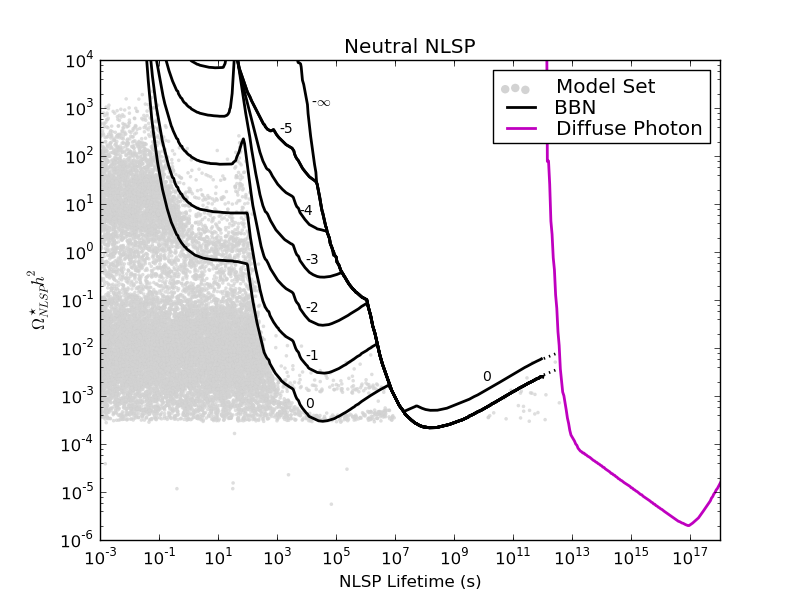}}
\vspace*{0.5cm}
\caption{Cosmological constraints on the density of decaying neutral NLSPs with a mass of 1 TeV and an electromagnetic branching fraction of 1. Black lines show BBN constraints \cite{Jedamzik:2006xz}, rescaled for our model set as described in the text. Lines are labeled by the logarithm of the hadronic branching fraction. The purple line is the limit derived from measurements of the diffuse photon flux. Constraints on lighter NLSPs are generally slightly stronger or equivalent to those shown here.}
\label{fig:neutconst}
\end{figure}

Figure~\ref{fig:neutconst} shows the constraints we obtain for a 1 TeV neutral NLSP having a negligible invisible width.  The black curves in this figure
represent the BBN bounds for various values of the hadronic branching fraction (logarithmically labeled).  Our pMSSM models are represented by the gray points, and each model lies
below the appropriate curve for its hadronic branching fraction.  Limits on lighter NLSPs are generally equivalent to or slightly stronger than those shown as discussed above.
The effect of phase-space suppression can be seen in the figure, where the limit for lifetimes $> 10^7$ seconds would be nearly flat for unsuppressed decays, but becomes significantly less severe as the gravitino carries away an increasing fraction of the available energy.

The BBN constraints on long-lived particles described above include only the effects of electromagnetically and hadronically-interacting decay products of the NLSP. However, long-lived particles can have additional contributions to BBN. NLSPs carrying strong or electromagnetic charge may interact directly with nucleons, affecting their reactions and stability. Additionally, neutrinos resulting from NLSP decays may scatter on background leptons, and the resulting final states can affect BBN.


Negatively charged particles living longer than $1.2 \times 10^3$ seconds can form bound states with $^4He$ as well as with protons and other nuclei~\cite{Kohri:2009mi}. This formation of bound states with charged particles can increase the probability that their decay products will interact with the nuclei due to proximity~\cite{Kaplinghat:2006qr}. Bound state formation can also alter the rates for nuclear reactions, for example by lowering the coulomb barrier~\cite{Kohri:2006wk}. A comprehensive study of the constraints on long-lived charged particles was performed in~\cite{Jedamzik:2007qk}, which includes the effects of both bound state formation and energetic particle injection. In this work, constraints are given for particle masses of 100 GeV and 1 TeV for a variety of hadronic branching fractions. Since charged particles affect BBN in more complicated ways than neutral ones, we do not attempt to guess a mass-dependence for these constraints and rely instead on a linear interpolation between the constraints for 100 GeV and 1 TeV particles. Although this interpolation may slightly overestimate the strength of the bounds in regions where the hadronic energy injection is dominant, we expect this difference to be at most a factor of 2 due to the larger number of baryons produced by a generic 1 TeV particle compared with our NLSPs as discussed above.  This effect is well compensated by the factor of 2 uncertainty in our calculation, and hence we do not expect to be removing cosmologically viable NLSPs from our model set.

\begin{figure}
\centering
\subfloat{\includegraphics[width=6.5in]{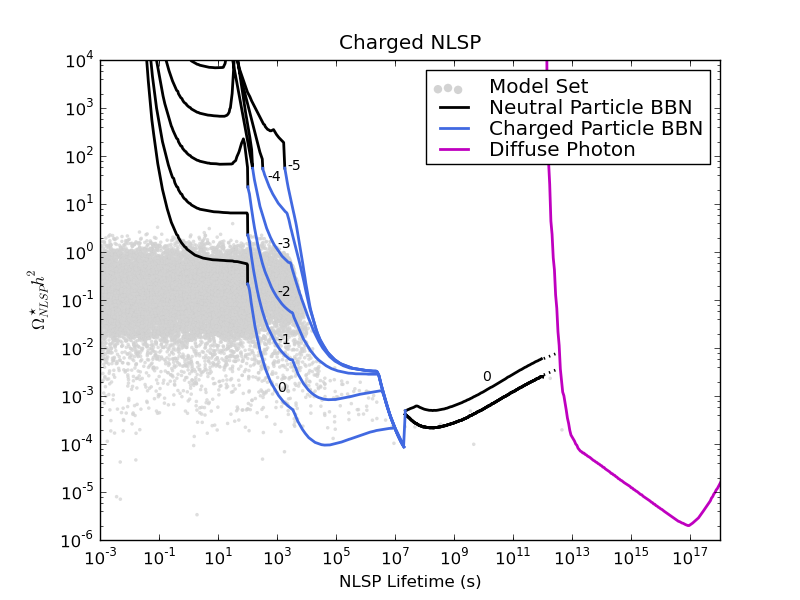}}
\vspace*{0.5cm}
\caption{Cosmological constraints on the density of decaying charged NLSPs with a mass of 1 TeV. Black lines show BBN constraints on neutral NLSPs~\cite{Jedamzik:2006xz}, which are also applicable to charged particles in these regions. Blue lines show BBN constraints on charged NLSPs~\cite{Jedamzik:2007qk}, and are stronger than or equivalent to the neutral particle constraints for NLSP lifetimes in the approximate range $10^{2}-10^{7}$ s. The discontinuity at $2 \times 10^7$ seconds results from rescaling the neutral constraints to include phase-space suppression, and has a minimal effect on the model set. Lines are labeled by the logarithm of the hadronic branching fraction. The purple line is the limit derived from measurements of the diffuse photon flux, which is identical to the limit on neutral particles. Constraints on lighter NLSPs are generally slightly stronger or equivalent to those shown here.}
\label{fig:chrgconst}
\end{figure}

Translating these restrictions to our pMSSM model set yields the
constraints on charged NLSPs displayed in Fig.~\ref{fig:chrgconst} as denoted by the light blue curves.  We see that these considerations place the strongest bounds on
the NLSP density for lifetimes roughly in the range $10^{2}-10^{7}$ s. For shorter or longer lifetimes, energetic particle injection provides the dominant constraint, so the exclusion contours are identical to those for a neutral NLSP.

\begin{figure}
\centering
\subfloat{\includegraphics[width=6.5in]{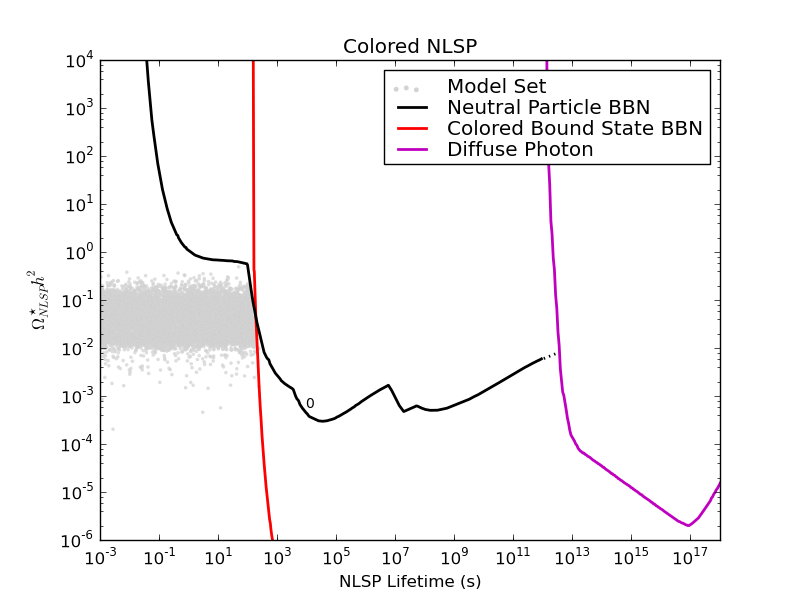}}
\vspace*{0.5cm}
\caption{Cosmological constraints on the density of decaying colored NLSPs with a mass of 1 TeV. The black line is the BBN limit on a neutral particle with a hadronic branching fraction of 100\%~\cite{Jedamzik:2006xz}. The red line shows the BBN bound on stable colored particles~\cite{Kusakabe:2009jt}. The purple line is the limit derived from measurements of the diffuse photon flux, which is identical to the limit on neutral particles. Constraints on lighter NLSPs are generally slightly stronger or equivalent to those shown here.}
\label{fig:colrconst}
\end{figure}

Heavy hadrons resulting from the hadronization of colored NLSPs can also form bound states with nuclei, strongly affecting BBN~\cite{Kusakabe:2009jt}. Since the heavy hadrons are strongly interacting, they form such bound states early, after t $\sim$ 10 s. These hadronic bound states can affect nuclear reaction rates by altering the coulomb barrier (if charged), modifying reduced masses, and changing reaction Q-values. In~\cite{Kusakabe:2009jt}, BBN constraints are derived for a neutral heavy hadron in this scenario. Due to the complicated nature of strong interactions, many of the relevant reaction rates are estimates, and the theoretical uncertainty of the results may be significant. There is also considerable theoretical uncertainty in the relic density of metastable colored particles as discussed above. Figure~\ref{fig:colrconst} compares the constraints on colored NLSPs resulting from hadronic bound state formation to those derived from hadronic energy injection. We see that in both cases the constraints become very strong for lifetimes of $\sim$ 200 s, when $^4He$ first becomes abundant, indicating that the maximum lifetime for metastable colored particles is relatively robust despite the theoretical uncertainties involved.

\begin{figure}
\centering
\subfloat{\includegraphics[width=3.2in]{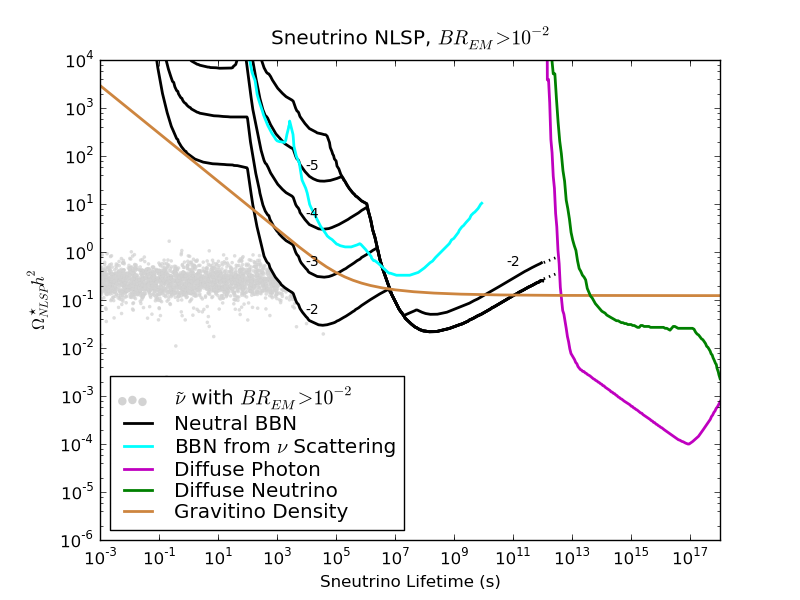}} ~
\subfloat{\includegraphics[width=3.2in]{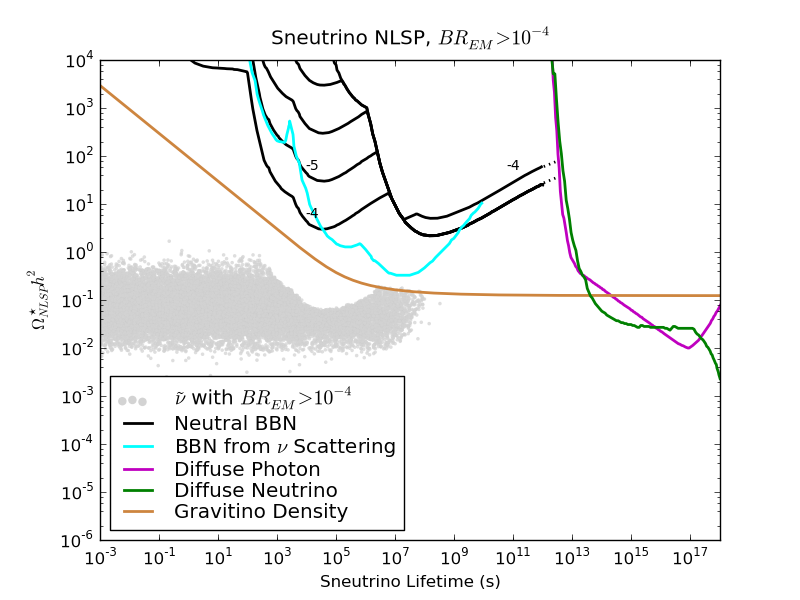}} \\
\subfloat{\includegraphics[width=3.2in]{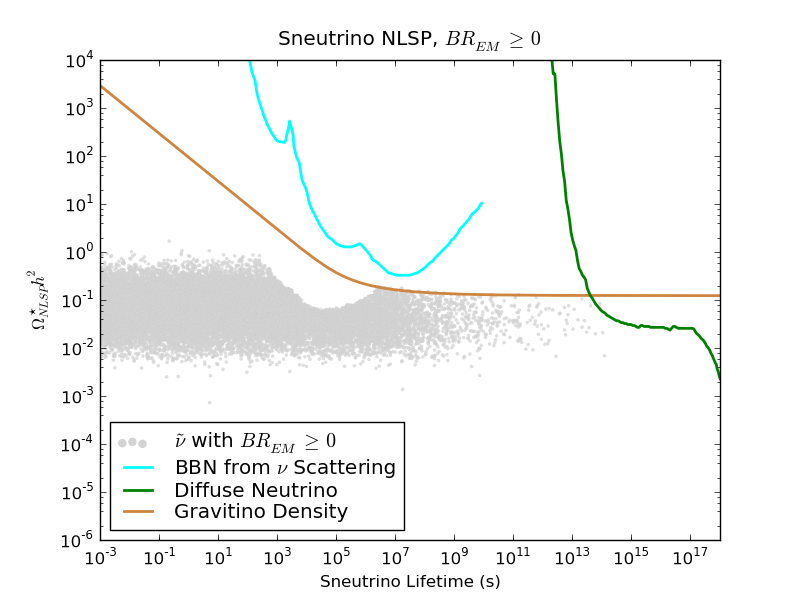}}
\vspace*{0.5cm}
\caption{Cosmological constraints on the density of decaying sneutrino NLSPs with a mass of 1 TeV. Black lines show BBN constraints on neutral particles~\cite{Jedamzik:2006xz} rescaled by the electromagnetic branching fraction of the sneutrino. The cyan line corresponds to the limit on neutrino injection during BBN~\cite{Kanzaki:2007pd}. The brown line represents the bound from overproduction of gravitinos, using the relation between the gravitino mass and sneutrino lifetime. The purple line is the limit derived from measurements of the diffuse photon flux, also rescaled by the electromagnetic branching fraction. The green line represents the bound from measurements of the diffuse neutrino flux, rescaled to reflect the energy of the injected neutrino as described in the text. Constraints on lighter sneutrinos are generally slightly stronger or equivalent to those shown here, although the scaling of the diffuse neutrino constraint is somewhat more complicated.}
\label{fig:sntrconst}
\end{figure}

The constraints described for neutral particles in~\cite{Jedamzik:2006xz} assume that the dominant decay products of the NLSP are hadrons, photons, or electrons, with neutrinos playing an insignificant role apart from carrying away some of the energy of the hadronic showers. Sneutrino NLSPs, on the other hand, decay dominantly to neutrinos and gravitinos, with only a small fraction of decays resulting in electromagnetically or hadronically interacting particles. However, high-energy neutrinos produced in NLSP decays can scatter off background leptons to produce charged leptons and mesons, which then affect BBN in the manner described above for neutral particles~\cite{Kanzaki:2007pd}. Despite the weak nature of neutrino interactions, constraints from neutrino scattering can dominate over those resulting from electromagnetic and hadronic decay products in cases where the sneutrino electromagnetic branching fraction is extremely small. Figure~\ref{fig:sntrconst} shows the bounds on sneutrinos with electromagnetic branching fractions of $10^{-2}$, $10^{-4}$, and $0$. For comparison, the restrictions from both neutrino scattering and from electromagnetic decays of the sneutrino are presented.  We see
that the strongest BBN constraints result from neutrino scattering processes once the electromagnetic branching fraction is small enough, $\lsim 0.1\%$, as expected.
We note that in our scan, sneutrinos with lifetimes in the range where neutrino scattering is important tend to have a small mass splitting with the gravitino. As a result, the upper bound on the gravitino relic density derived above is more stringent, as can be seen in the figure.

\subsection{Diffuse Photon Spectrum}
 
Particles decaying after recombination ($\tau_{NLSP} > 10^{13}$ s) can produce high energy $\gamma$-rays that would be seen as an excess in the diffuse photon spectrum~\cite{Kanzaki:2007pd}. These photons may be produced both directly in NLSP decays and by the scattering of leptonic and baryonic decay products with background particles. The resulting $\gamma$-rays are then attenuated through scattering with background photons. In~\cite{Kribs:1996ac}, measurements of the diffuse photon background from COMPTEL and EGRET are used to constrain the density of particles decaying electromagnetically at late times.  Here, it is found that the limit is nearly independent of the
injection energy for relatively early decays, and becomes stronger with decreasing photon energy at later injection times.
NLSPs with a long-enough lifetime to be affected by diffuse constraints have a small mass splitting with the gravitino, $\Delta M \lesssim 10$ GeV, and a correspondingly low decay product energy. We therefore use the lowest energy constraint available, corresponding to 25 GeV photons, and expect this to be a conservative limit.\footnote{Note that the analysis in~\cite{Chen:2003gz} indicates that no qualitative changes in the constraints occur for the energy range relevant to our model set.} The diffuse photon constraints used in our analysis are displayed in Figs.~\ref{fig:neutconst}-\ref{fig:sntrconst}, which show that they become strong for NLSP lifetimes longer than $10^{13}$ seconds. 
While stronger constraints than those in~\cite{Kribs:1996ac} could potentially result from a similar analysis of Fermi data, the impact on our model set is expected to be negligible. The reason is that non-sneutrino NLSPs are already strongly excluded in this region, while sneutrino NLSPs with lifetimes in the relevant range tend to have extremely suppressed decays to visible energy, with electromagnetic branching fractions of order $10^{-8}$ or smaller, reducing their diffuse photon signals far below the sensitivity of Fermi. 
 
\subsection{Diffuse Neutrino Spectrum}
 
Neutrinos resulting from the decay of extremely long-lived sneutrinos can lead to a detectable excess in the diffuse neutrino flux.  Measurements of atmospheric neutrinos from Super-Kamiokande and AMANDA and the relic supernova $\bar{\nu_e}$ flux from Super-Kamiokande can be used to calculate constraints on late-decaying particles that decay dominantly to neutrinos~\cite{Kanzaki:2007pd}. In this reference, constraints are given for a range of initial neutrino energies, starting at 100 GeV. Since extremely long-lived sneutrinos in our model set decay to lower energy neutrinos due to the small sneutrino-gravitino mass splitting (typically, $\Delta m \lesssim 5$ GeV), we must extrapolate these bounds to lower energies. We do this by rescaling the computed limits on 100 GeV neutrino injection. For example, we consider a 10 GeV neutrino produced at a redshift of $z = 100$ as equivalent to a 100 GeV neutrino produced at $z = 1000$. The scale factor at the effective injection time is calculated from the scale factor at the true injection time through the relation $a_{eff} = \frac{E_{\nu}}{100 \: GeV} a_{true}$, and is given as a function of time in~\cite{Sazhin:2011ze}, using parameter values from WMAP 7~\cite{Komatsu:2010fb}.  Although this approximation is imperfect, since sneutrinos with a given lifetime decay at a range of redshifts, the consequence of the approximation for our model set is insignificant.  The restrictions placed on our pMSSM model set from this constraint can be observed in Fig.~\ref{fig:sntrconst}. 

\subsection{Other Cosmological Observables}

Several other cosmological observables may be affected by the presence of late-decaying particles. These include the CMB chemical potential and Compton y-parameter, the correlation between E-type polarization and temperature in the CMB, and measurements of large-scale structure. We analyzed these effects and determined that they are either inapplicable to our range of NLSP masses and lifetimes or yield degenerate constraints with those described above.

\subsection{Results of Gravitino LSP Model Generation}

After performing the scan described in Section \ref{sec:modgen-grav} and applying the cosmological constraints described above, we generate a gravitino LSP model set consisting of 229,303 models consistent with all known experiments other than the LHC, for which only preliminary constraints, such as limits on pair production of stable particles, have been applied. Future work will explore the full LHC phenomenology of the gravitino model set~\cite{future}. Nearly all of the gravitino LSP models require modifications of the standard cosmology to prevent gravitinos from being overproduced, but these modifications could generally occur at a high temperature and are unimportant for the weak-scale phenomenology that we study below. We can now analyze the properties of these models and compare them with the neutralino LSP model set.

\subsection{Summary of Cosmological impacts on the Gravitino LSP Model Set}

The importance of including a gravitino in our model set can be understood by analyzing the characteristics of the surviving models. We will see that both cosmological considerations and stable particle searches remove a significant portion of the available parameter space, and are therefore highly relevant for this scenario.

\begin{figure}
\centering
\subfloat{\includegraphics[width=6.5in]{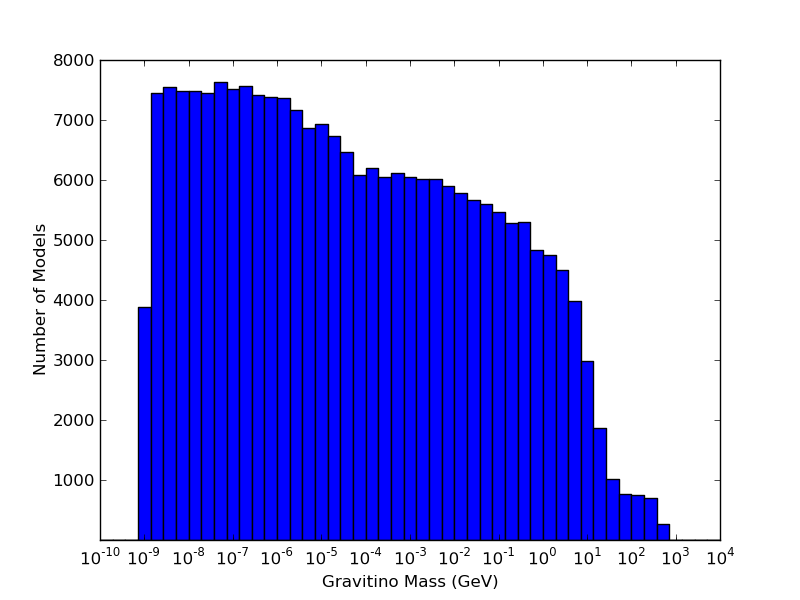}}
\vspace*{0.5cm}
\caption{Distribution of gravitino masses in the gravitino pMSSM.}
\label{fig:gravmass}
\end{figure}

The striking importance of the global constraints on the gravitino mass can be seen in Fig.~\ref{fig:gravmass}, which shows the number of surviving models with a given gravitino mass. Models with very light gravitinos do not contain stable particles, so their viability is not sensitive to the particular value of the gravitino mass. As the gravitino becomes increasingly massive, more and more models contain stable particles and the number of surviving models decreases significantly, resulting in the step at
masses of $\sim 10^{-5}$ GeV. A second decrease in the number of viable models occurs for gravitinos heavier than 10 MeV, where cosmological constraints, particularly BBN, become increasingly important.

\begin{figure}
\centering
\subfloat{\includegraphics[width=6.5in]{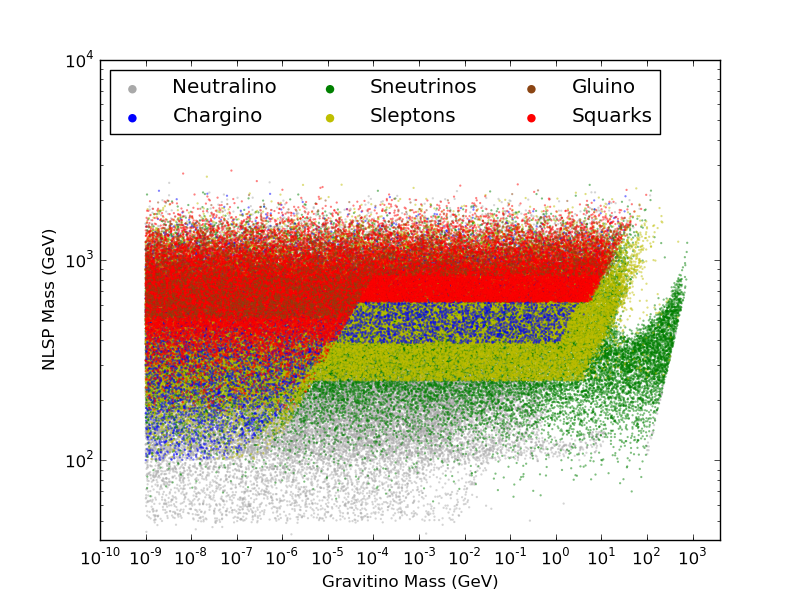}}
\vspace*{0.5cm}
\caption{Viable models in the NLSP and gravitino mass plane, where the color codes represent the NLSP types as labeled. The effects of model-independent limits on stable charged particles can be observed as sloped and horizontal lines for smaller gravitino masses, while cosmological constraints produce the structure seen in the right part of the graph for gravitinos heavier than $\sim$ 1 GeV.}
\label{fig:gravnlsp}
\end{figure}

A useful way to visualize the available parameter space is to study the relationship between the NLSP and gravitino masses, as shown in Fig.~\ref{fig:gravnlsp}. Here, we again see the strong effect of the constraints on our gravitino model set. In particular, the importance of the NLSP identity for both collider and cosmological constraints is clearly evident. The effects of the collider stable particle searches result in the sharp sloped line starting at gravitino masses of $\sim 10^{-7} - 10^{-6}$ GeV as well as the
sharp horizontal line.  The slope of the stable particle bound originates from the strong dependence of the NLSP width on its mass, which goes roughly as $\frac{M_{NLSP}^5}{M_{\tilde{G}}^2}$, and then flattens out once the NLSP mass reaches the kinematic limits of the LHC searches. We expect the stable particle limits to become even stronger after more LHC analyses are completed; these will extend the limits on chargino and slepton NLSPs by including searches for the indirect production of stable particles through a decay chain, as well as for looking for particles that decay within the detector.  The chunk of parameter space that is removed for heavier gravitinos (starting at masses of $\gsim 1$ GeV, corresponding to typical NLSP lifetimes $\gsim 100$ s) is due to the constraints
from BBN. 

\begin{figure}
\centering
\subfloat{\includegraphics[width=6.5in]{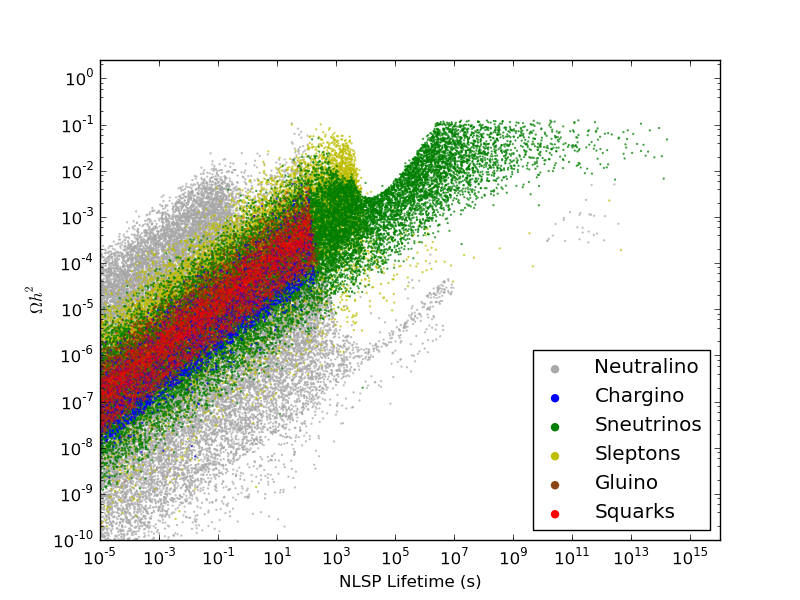}}
\vspace*{0.5cm}
\caption{The non-thermal relic density of gravitinos produced by different NLSP types. Note that sneutrino decays can saturate the WMAP bound, while other NLSP types are generally prevented from doing so by BBN.}
\label{fig:nlsplifegravomega}
\end{figure}

The non-thermal relic density of the gravitino is shown in Fig.~\ref{fig:nlsplifegravomega}, which reveals the cosmological effects of the gravitino. In some cases, we see that the gravitino relic density can saturate the WMAP bound through non-thermal production of gravitinos in NLSP decays. Since the gravitino relic density is tied to the weak-scale relic density of the NLSP, this mechanism preserves the WIMP miracle without requiring dark matter to be a WIMP. From the figure, we see that this saturation is easily accomplished when the sneutrino is the NLSP, since sneutrino decays have a comparatively small effect on BBN. Saturation of the relic density is also possible for neutralinos and sleptons, but in this case the surviving models are very close to the exclusion contours and are therefore likely to be in significant tension with data.

Of all of the cosmological considerations discussed above, BBN excludes by far the most models. The diffuse photon and neutrino spectra exclude very few of our models because they are relevant only for extremely long NLSP lifetimes; this can only be achieved by a small splitting between the NLSP and gravitino, which is statistically unlikely. Consistency with the WMAP upper bound on the gravitino relic density only eliminates models with a sneutrino NLSP, and hence only affects a small fraction of the total model set. An improvement in the BBN limits would significantly enhance the constraints on sleptons and sneutrinos, but would be unlikely to have much effect on the other NLSP types (which are already strongly constrained by BBN). Interestingly, Fig.~\ref{fig:sntrconst} shows that BBN constraints which include the effects of neutrino injection are close to excluding some of the interesting region for non-thermal gravitino production. Significant improvements in the limits on relativistic energy injection, which are derived from CMB parameters and measurements of large-scale structure~\cite{Kanzaki:2007pd}, could also help constrain late-decaying sneutrinos.


We have seen here how constraints from collider studies and cosmology constrain the properties of the NLSP and gravitino. We now discuss the resulting differences in the properties of the neutralino and gravitino model sets with respect to a variety of observables.


\section{Comparison of Neutralino and Gravitino Model Sets}
\label{sec:compare}

Superficially, since the scan ranges of the 19 soft parameters (excluding the gravitino mass) are identical in both the neutralino and gravitino LSP model sets, 
one might expect the general characteristics of the two sets to be very similar. While this expectation holds to some degree, the details are 
somewhat more complex. As discussed in the previous section, there are some significant differences in the astrophysical and cosmological constraints on the 
two model sets. While there are obvious differences in the two LSP properties, their direct impact on LHC searches is generally relatively minor. However, there are some important distinctions which can indirectly influence LHC searches which we now discuss. 

\begin{figure}
\centering
\subfloat{\includegraphics[height=3.5in]{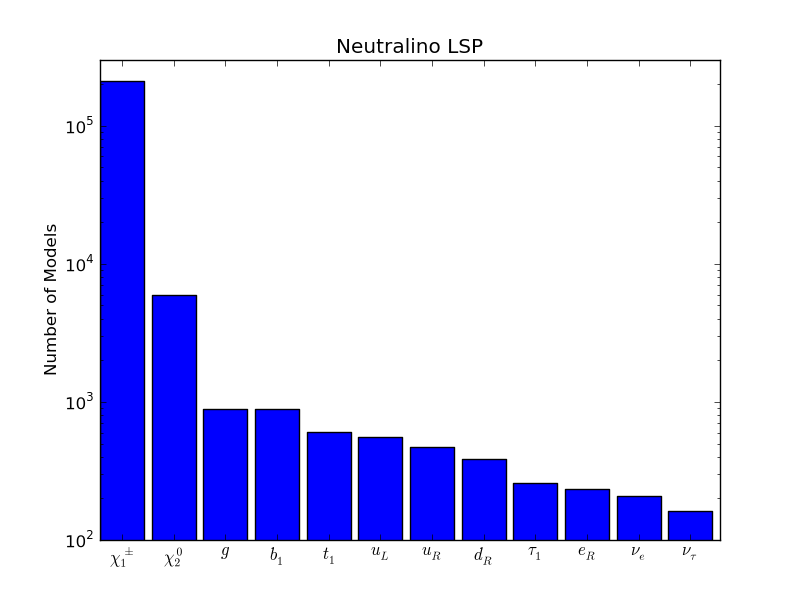}} \\
\subfloat{\includegraphics[height=3.5in]{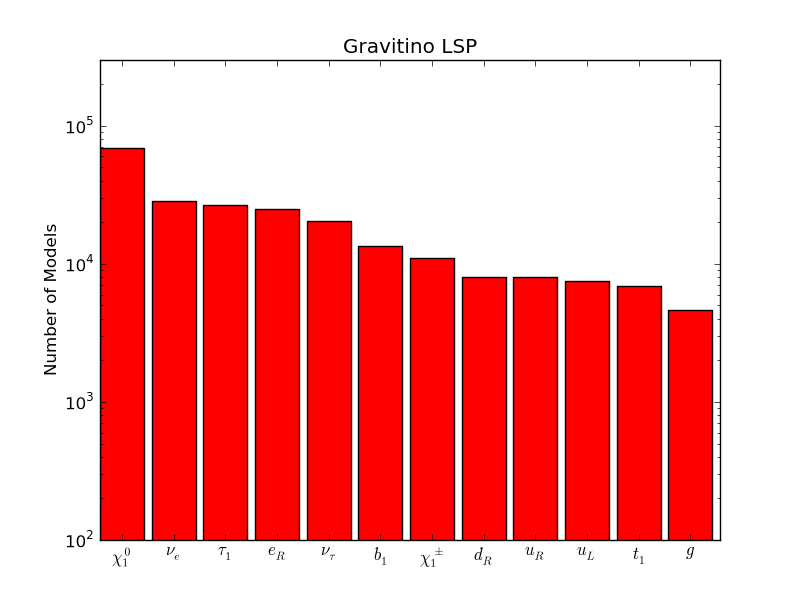}}
\vspace*{0.5cm}
\caption{The frequency of the various NLSP candidates in the neutralino (top) and gravitino (bottom) LSP model sets.  Note that the ordering of the NLSP type on the horizontal axis differs 
between the two panels.}
\label{fig:nlspid}
\end{figure}

First, the identity of the NLSP, which occasionally has a critical effect on the LHC signatures of a model, is very sensitive to the choice of the LSP; this can be seen in Fig.~\ref{fig:nlspid}. When the neutralino is the LSP, the NLSP is most frequently a chargino or a second neutralino. In this case, $\tilde\chi_1^\pm$ and $\tilde\chi_2^0$ are preferred NLSP states because the LSP is nearly always a relatively pure wino or Higgsino, as shown in Table~\ref{tab:neutcomp}.  Generally, this implies the existence of a chargino with a mass very similar to that of the NLSP. When the LSP is a nearly pure Higgsino, a second neutralino also has a small mass splitting with the LSP and is therefore likely to be the NLSP. The top panel in Fig.~\ref{fig:nlspid} demonstrates the strong preference for neutralino and chargino NLSPs in this model set. These two possibilities alone account for 
$\sim 89\%$ of the models in the neutralino model set. If the LSP is mostly a bino, it is not required to have a similar mass to any other sparticle, and we would expect the NLSP identity to be essentially arbitrary, as the various mass parameters are then at least approximately uncorrelated. In addition, the requirement that the LSP relic abundance is not so large as to overclose the universe places further constraints on the NLSP. Since heavy bino LSPs usually have an excessive relic density, some co-annihilation is often necessary to achieve consistency with the WMAP observation.   Models with NLSPs which co-annihilate efficiently are therefore more likely to pass the constraints. Looking at Fig.~\ref{fig:nlspid}, we see that the most common NLSP types after charginos and neutralinos are strongly-interacting sparticles, followed by charged and finally neutral sparticles, as we would expect based on their respective abilities to co-annihilate.

\begin{table}
\begin{center}
\begin{tabular}{|c|c|c|c|}
\hline
Lightest Neutralino & Definition & $\tilde{\chi}_1^0$ LSP & $\tilde{G}$ LSP \\
\hline
\hline
Bino & $|N_{11}|^2 > 0.95$ & 0.024 & 0.313 \\
Mostly Bino & $0.80 < |N_{11}|^2 < 0.95$ & 0.002 & 0.012 \\
Wino & $|N_{12}|^2 > 0.95$ & 0.546 & 0.296 \\
Mostly Wino & $0.80 < |N_{12}|^2 < 0.95$ & 0.022 & 0.019 \\
Higgsino & $|N_{13}|^2 + |N_{14}|^2 > 0.95$ & 0.340 & 0.296 \\
Mostly Higgsino & $0.80 < |N_{13}|^2 + |N_{14}|^2 < 0.95$ & 0.029 & 0.029 \\
All other models & $|N_{11}|^2, |N_{12}|^2, |N_{13}|^2 + |N_{14}|^2 < 0.80$ & 0.036 & 0.035 \\
\hline
\end{tabular}
\caption{The fraction of models which have the lightest neutralino with various bino, wino or Higgsino purities in both the neutralino and gravitino LSP model sets.}
\label{tab:neutcomp}
\end{center}
\end{table}

When the gravitino is the LSP, the identity of the NLSP is essentially arbitrary (aside from any possible bias imposed by the scan ranges) before we apply the set of experimental and cosmological constraints described in sections 2 and 3. As discussed at length above, the NLSP identity is strongly affected by both stable particle searches at colliders and BBN constraints. Stable particle limits are determined by the pair-production cross-section and interaction strength of a sparticle, so the limits are strongest for gluinos and squarks, intermediate in size for charginos and sleptons, and non-existent for neutral particles. Since BBN constrains electromagnetic and hadronic energy injection, it tends to restrict the same NLSP types that are favored by stable particle searches. However, there are a couple of differences between the two. First, charginos, which always have an $\cal O$(1) hadronic branching fraction, thus contributes as much as colored sparticles to BBN and have similar bounds. Second, neutralinos nearly always have charged decay products despite being neutral themselves, and therefore face much stronger BBN constraints than sneutrinos. The resulting distribution of NLSP types in the gravitino model set is a combination of these stable particle and BBN preferences. In Fig.~\ref{fig:nlspid}, we see that gluinos, squarks, and charginos are the NLSP less often on average, while neutralinos and sneutrinos are more frequently the NLSP. The excessive prevalence of a $\tilde\chi_1^0$ NLSP results from statistics, since any of the four neutralinos may be the NLSP.

\begin{figure}
\centering
\subfloat{\includegraphics[width=4.5in]{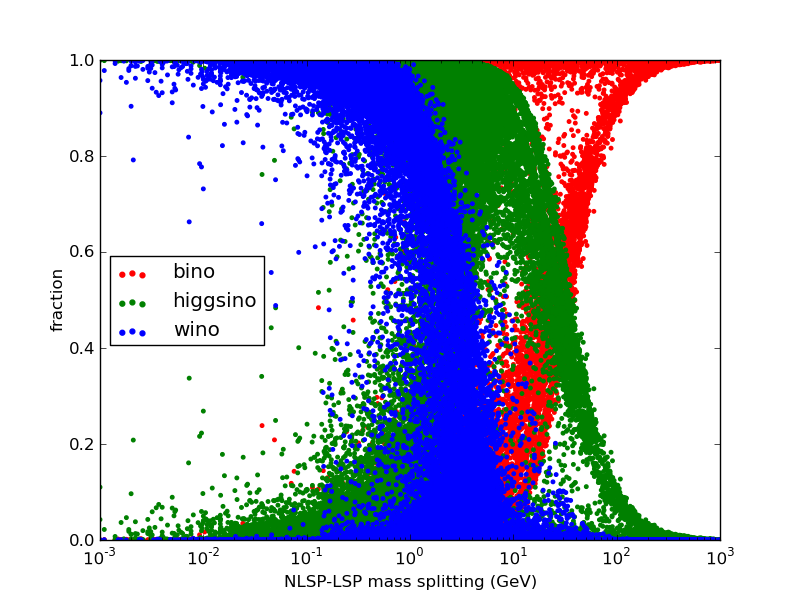}}
\vspace*{0.5cm}
\caption{The electroweak composition of the neutralino LSP as a function of its mass splitting with the NLSP. The LSP's electroweak content is shown as 3 points 
         for each model.} 
\label{fig:lspewk}
\end{figure}

If the LSP is the neutralino then its electroweak content and mass splitting with the NLSP are important characteristics that affect collider searches. Figure~\ref{fig:lspewk} 
displays the evolving bino, wino and Higgsino content of the lightest neutralino as a function of its mass splitting with the NLSP. Here, we see that for a very small 
mass splitting, below $\sim 1$ GeV, the neutralino is likely to be mostly wino-like. For somewhat larger mass splittings, of order a few to ten GeV, the neutralino is 
mostly Higgsino-like while for even larger splittings it is seen to be mostly bino. 

\begin{figure}
\centering
\subfloat{\includegraphics[width=6.5in]{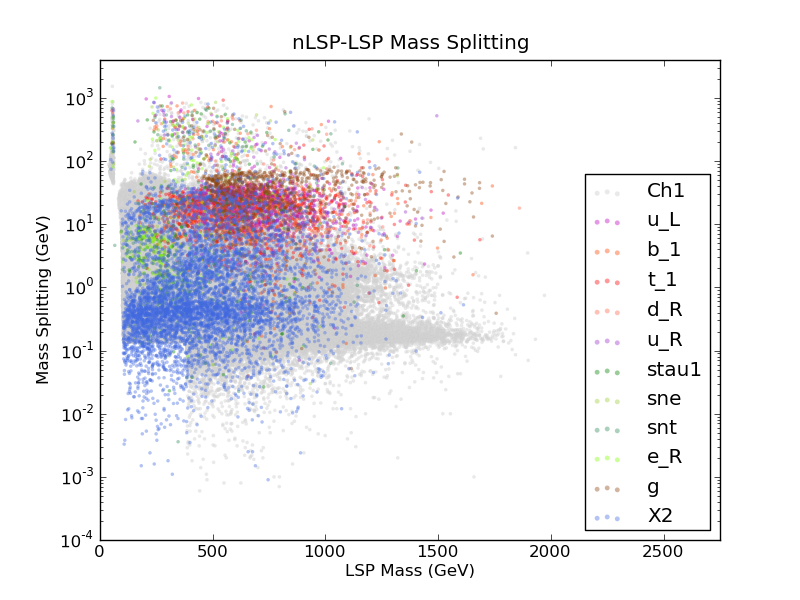}}
\vspace*{0.5cm}
\caption{The mass splitting between the NLSP and the neutralino LSP as a function of the LSP mass. The various possible identities of the NLSP are color-coded. 
         Note the box-shaped region in the lower left-corner, which is removed for the case of chargino NLSPs due to the stable particle searches described in the text.} 
\label{fig:lspsplit}
\end{figure}

Figure~\ref{fig:lspsplit} shows the mass splitting between the neutralino LSP and the various possible NLSP types as a function of the LSP mass, with the identity of 
the NLSP color-coded, in the neutralino model set. The range of NLSP-LSP mass splittings in this model set, from below $\sim 1$ MeV to greater than $\sim 1$ TeV, is impressively large. 
Note that (as expected) both chargino and second neutralino NLSPs tend to have relatively smaller mass splittings with the LSP than do any of the colored 
superpartners, while sleptons tend to have mass splittings which are intermediate between these two possibilities. As discussed above, colored sparticles and sleptons tend to have splittings in a range which allows for efficient co-annihilation when the NLSP is a heavy bino, although there is a small population of models for which this is not the case. Note that a boxed-shaped region has been cut out of the chargino distribution for small splittings ($\Delta m \lesssim 120$ MeV) and low LSP masses. This square region shows the effect of stable particle search constraints, which were discussed above. 

\begin{figure}
\centering
\subfloat{\includegraphics[height=3.5in]{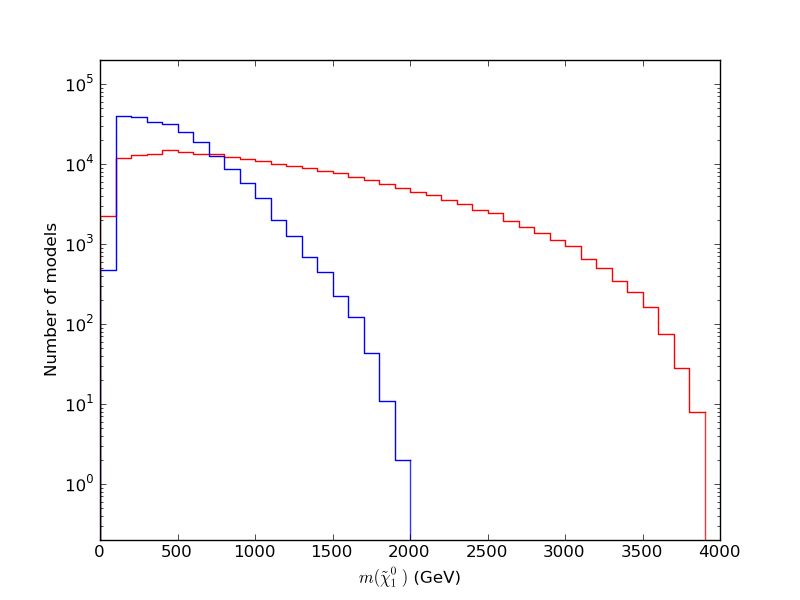}} \\
\subfloat{\includegraphics[height=3.5in]{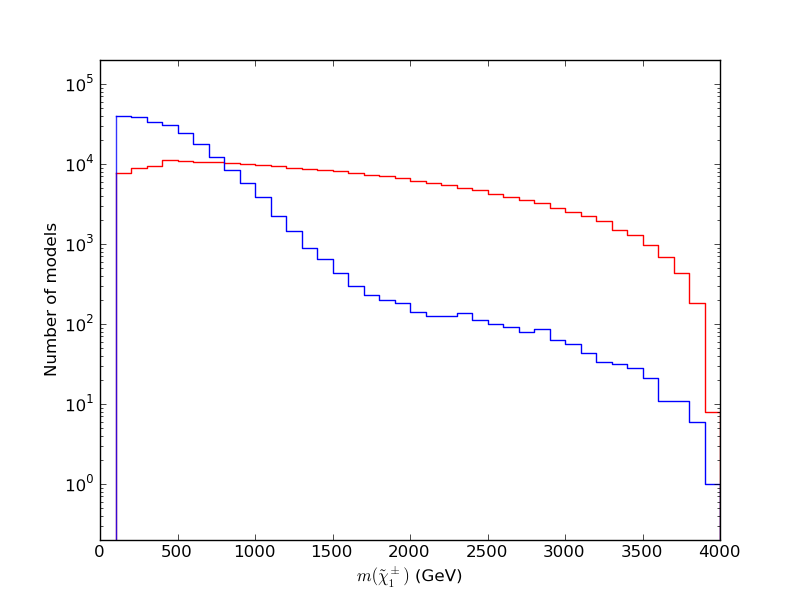}}
\vspace*{0.5cm}
\caption{The distribution of the lightest neutralino (top) and chargino (bottom) masses among the models in the neutralino LSP (blue) and gravitino (red) LSP sets.} 
\label{fig:neutcharspec}
\end{figure}

\begin{figure}
\centering
\subfloat{\includegraphics[height=3.5in]{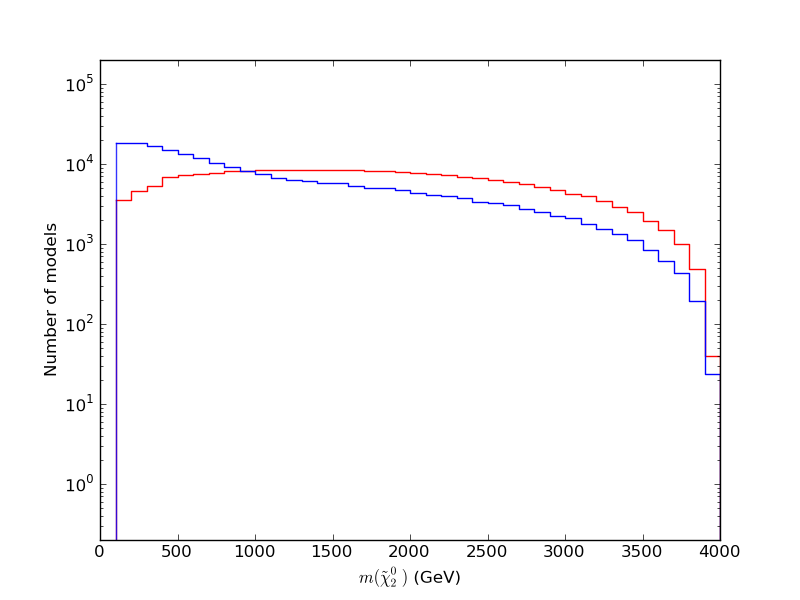}} \\
\subfloat{\includegraphics[height=3.5in]{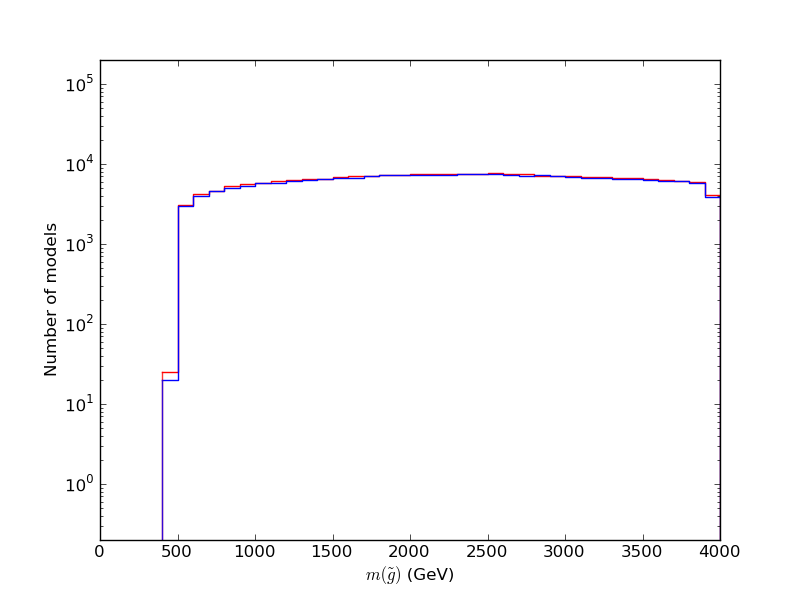}}
\vspace*{0.5cm}
\caption{The distribution of the second neutralino (top) and gluino (bottom) masses among the models in the neutralino LSP (blue) and gravitino (red) LSP sets.} 
\label{fig:spect2}
\end{figure}

\begin{figure}
\centering
\subfloat{\includegraphics[height=3.5in]{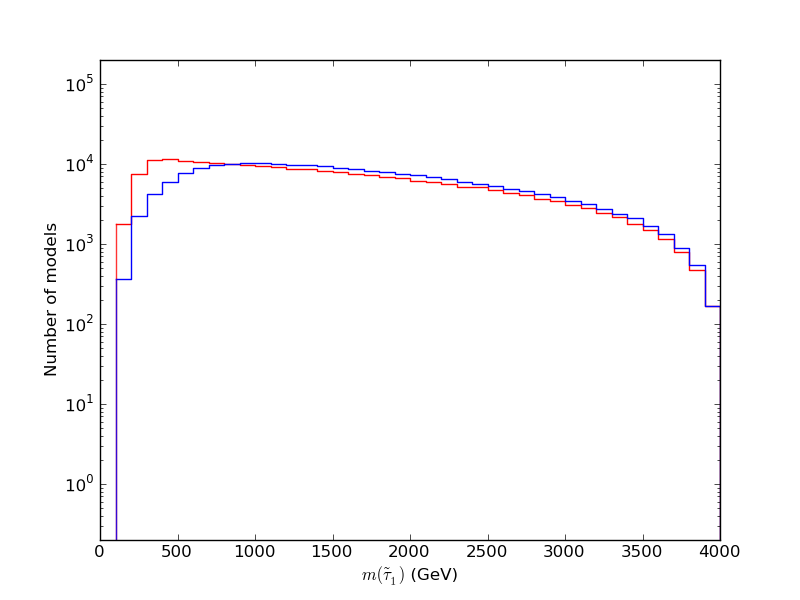}} \\
\subfloat{\includegraphics[height=3.5in]{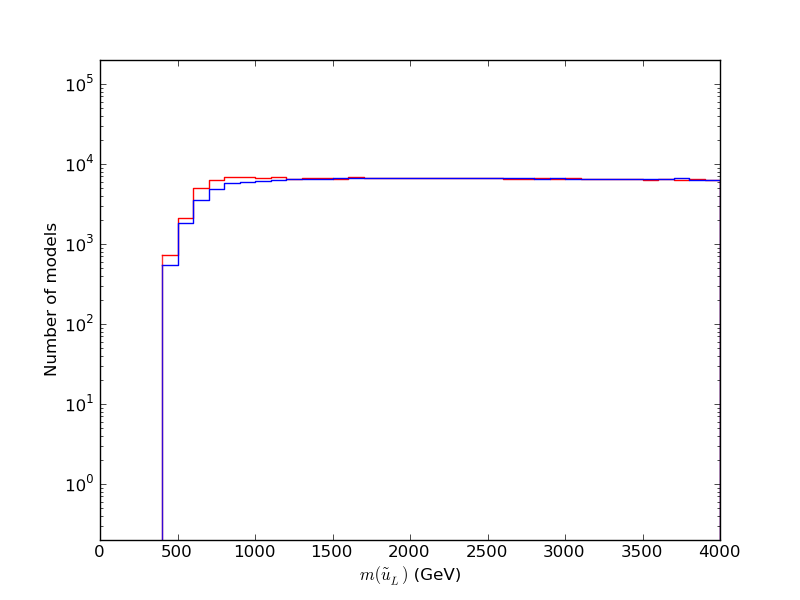}}
\vspace*{0.5cm}
\caption{The distribution of the lightest stau (top) and left-handed up-squark (bottom) masses among the models in the neutralino LSP (blue) and gravitino (red) 
          LSP sets.  }
\label{fig:spect3}
\end{figure}

The second most important consequence of choosing either a neutralino or gravitino as the LSP is its influence on the mass distributions of the other SUSY particles as well as, indirectly, that of the lightest Higgs scalar. While the gravitino is usually very light (well below $\sim 1$ GeV, say), the neutralino mass must be at least 
$>$ 10's of GeV due to our choice of scan ranges, and is frequently much heavier\footnote{The lower bound on the $\tilde\chi_1^0$ mass from collider and other data is approximately 7 GeV~\cite{Nakamura:2010zzi}.}. The physical masses of the sparticles in the gravitino model set can thus extend to somewhat lower values since they are not forced to be more massive than a neutralino LSP. Since most colored sparticles are already constrained to lie above $\sim 400$ GeV, however, the effect of this difference on their mass spectra is minor. For colorless states, which have lower bounds of $\sim 100$ GeV or less, and stops, which must only lie above $m_t$, the lower bound on their mass spectra resulting from being heavier than the LSP is much more significant, as we will see below. The lightest neutralino itself (and its frequently associated chargino and/or second neutralino) feel the strongest difference in their mass distributions, since they are no longer required to be light. The quantitative differences between the model sets can best be appreciated by considering Figs.~\ref{fig:neutcharspec},~\ref{fig:spect2} and~\ref{fig:spect3} which compare the distributions for some of the most relevant sparticles in the neutralino and gravitino model sets.    

In Fig.~\ref{fig:neutcharspec} we see the mass distributions for the lightest neutralino and chargino for both model sets. Figure~\ref{fig:spect2} shows the corresponding
distributions for the second neutralino and the gluino, respectively. The neutralino mass spectra reflect our discussion above; the lightest neutralino is no longer required to be very light in the gravitino scenario since it need not even be the NLSP, resulting in a much flatter distribution. Light neutralinos, with masses below 100 GeV, are also slightly more common in the gravitino model set due to the increased abundance of bino-like $\tilde\chi_1^0$ states, which can be light without having an associated chargino NLSP that is excluded by searches at LEP. In the neutralino model set, the prevalence of light neutralinos extends to a preference for light charginos and second neutralinos, as discussed previously. We therefore see this pattern, with light neutralinos and charginos dominating in the neutralino model set and a relatively flat distribution in the gravitino case. In the gravitino set, the chargino mass distribution is also affected by BBN; charginos which are 
light, \eg, the NLSP, are somewhat disfavored by the BBN constraints discussed above. This further promotes a flat mass distribution for the  $\tilde \chi_1^\pm$ in this model set.  For the gluino, which is always constrained to lie above 400 GeV while the LSP is generally significantly lighter in 
both model sets, Fig.~\ref{fig:spect2} shows that there is very little difference between the mass distributions in the two model samples.

In Fig.~\ref{fig:spect3} we present the mass distributions for both $\tilde \tau_1$ (taken as an example of a typical slepton) and $\tilde u_L$ (taken as an example of a typical squark); we note that the 
corresponding mass distributions for the other sleptons and first and second generation squarks are quite similar. Since light squarks are constrained to 
have masses above 400 GeV in our scans, there is very little difference apparent in the mass distributions between the two model sets. Sleptons, however, 
can be as light as 100 GeV and thus will feel the effect of the different LSP identity somewhat more strongly, explaining the small difference in the low-mass region of the two distributions.

\begin{figure}
\centering
\subfloat{\includegraphics[height=3.5in]{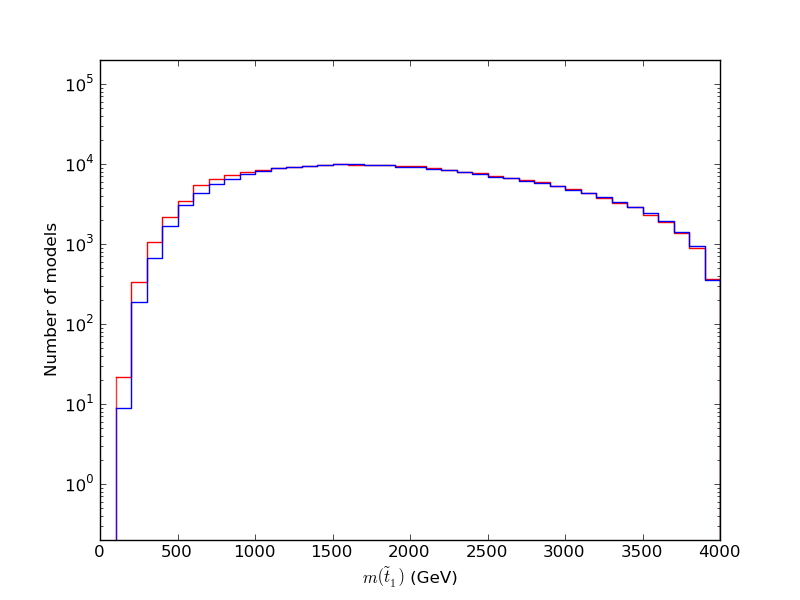}} \\
\subfloat{\includegraphics[height=3.5in]{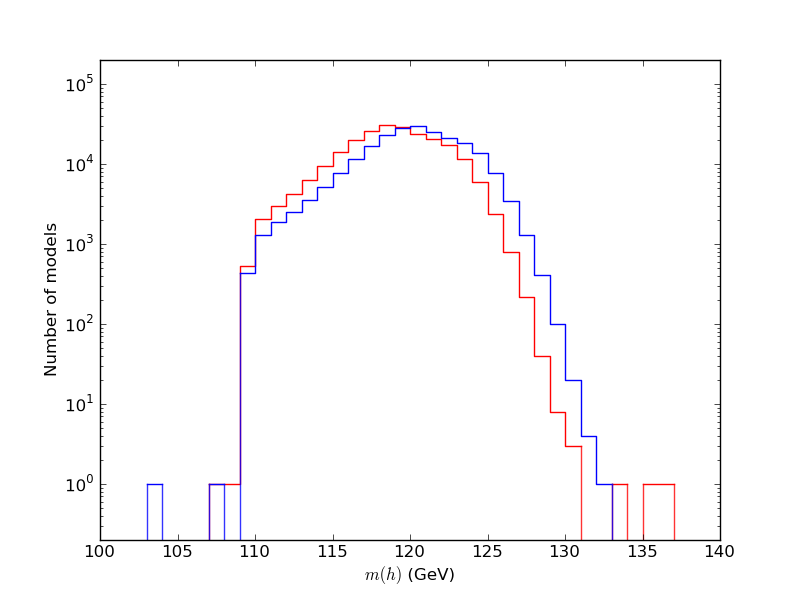}}
\vspace*{0.5cm}
\caption{The distribution of the lightest stop (top) and the lightest scalar Higgs (bottom) masses among the models in the neutralino LSP (blue) and gravitino (red) 
          LSP sets.}
\label{fig:spect4}
\end{figure}

In Fig.~\ref{fig:spect4} we see the mass distribution for $\tilde t_1$ which is constrained to lie above the top mass in the scan. Not surprisingly, the stop (which can be lighter than the first or second generation squarks) shows a greater difference between the neutralino and gravitino model sets than the first two generations of squarks.  
The preference for slightly lighter stops in the gravitino model set compared to the neutralino case has some important implications. Since 
both sets produce very similar distributions for $A_t$, $M_A$ and $\tan \beta$, the lighter stops in the gravitino set shift the predicted mass distribution for the lightest CP-even Higgs boson, $h$, to lower masses. This can be observed in Fig.~\ref{fig:spect4}. This implies, for example, that it is more difficult to get a mass for $h$ in the interesting range of $\simeq 123-127$ GeV in the gravitino set than in the neutralino set. The gravitino set will consequently have higher values of fine-tuning. These observations, along with other aspects of pMSSM Higgs phenomenology, will be discussed elsewhere \cite{future}.    

\begin{figure}
\centering
\subfloat{\includegraphics[width=3.2in]{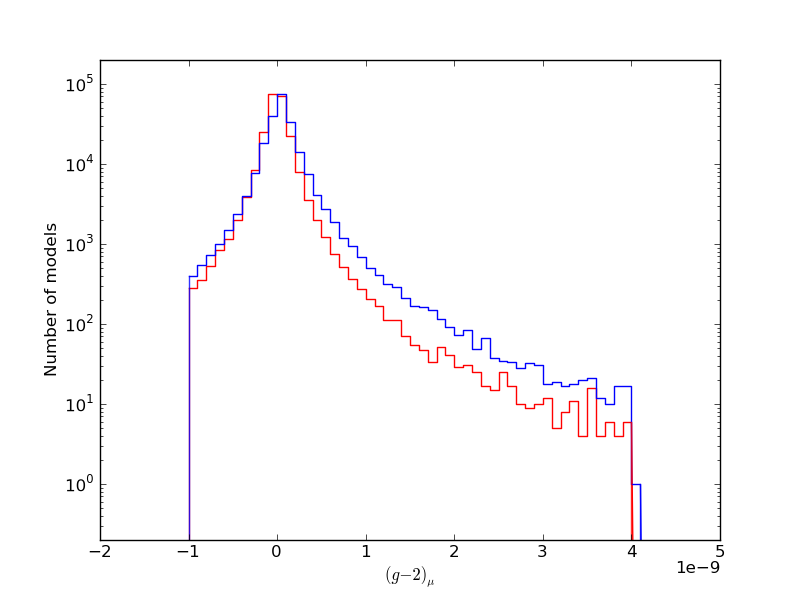}} ~
\subfloat{\includegraphics[width=3.2in]{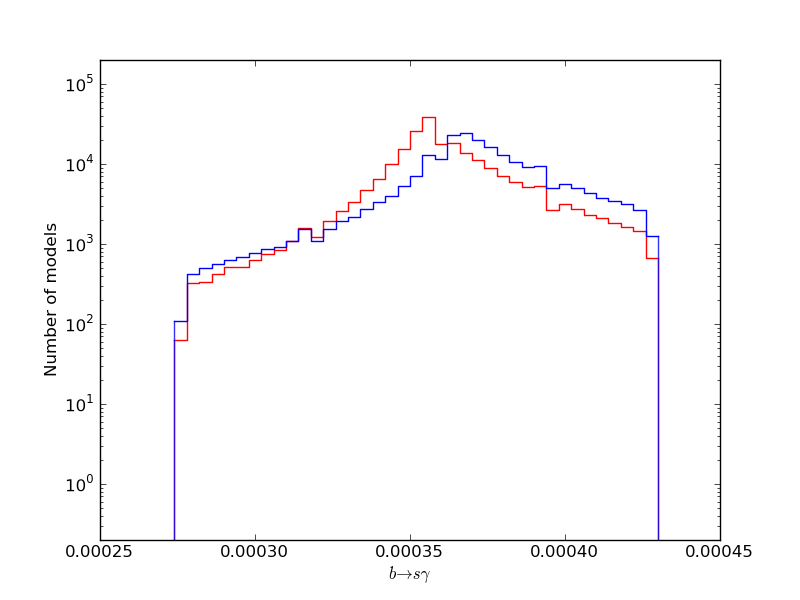}} \\
\subfloat{\includegraphics[width=3.2in]{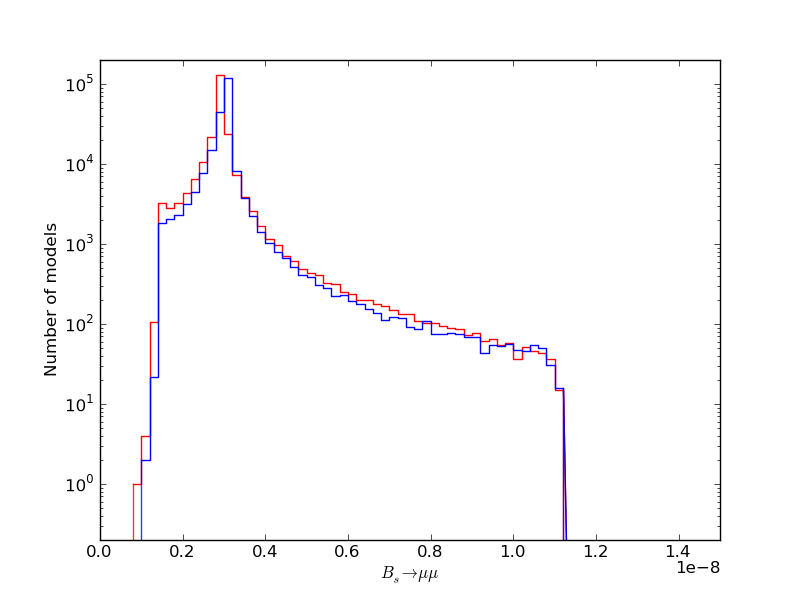}}
\vspace*{0.5cm}
\caption{The distribution for the predicted values of $g-2$ (upper left), $b\to s\gamma$ (upper right) and $B_s\to \mu^+\mu^-$ (bottom) in the neutralino LSP 
         (blue) and gravitino (red) LSP model sets.}
\label{fig:spect5}
\end{figure}

One may wonder if other observables are influenced significantly by the choice of LSP as a result of the small differences in sparticle mass distributions between the model sets. In Fig.~\ref{fig:spect5} 
we see the predicted distribution of values for the anomalous magnetic moment of the muon, as well as branching fractions of both $b\to s\gamma$ and $B_s\to \mu^+\mu^-$ for the two model sets. (Note 
that the ends of these distributions are cut off due to the finite scan ranges.) Some small differences in the predictions of the two sets are 
indeed observed but, at present, they are generally not very significant. The impact of the recent improvements to 
the upper bound on the branching fraction for $B_s\to \mu^+\mu^-$ by ATLAS~\cite{atlasbs}, CMS{\cite {cmsbs}} and LHCb{\cite {lhcbbs}} on our model sets will be discussed below.  

\section{LHC MET Searches}
\label{sec:met}

To test the potential of the LHC to discover or exclude our pMSSM models, we have simulated SUSY events for each model in the neutralino LSP set, leaving the gravitino LSP set to future work~\cite{future}. For each model, SOFTSUSY 3.1.7~\cite{Allanach:2001kg} was used to generate the spectrum and SUSY-HIT 1.3~\cite{Djouadi:2006bz} was used to calculate the decays for each sparticle, with the modifications noted in Section~\ref{sec:modgen}. Events were then generated for proton-proton collisions at 7 and 8 TeV using PYTHIA 6.4.26~\cite{Sjostrand:2006za} and PGS 4~\cite{PGS}. We considered all relevant SUSY processes, including associated and pair production of all $R$-odd particles. Events were also generated for both single charged Higgs production and for all two-Higgs final states that can proceed through an s-channel gauge boson. All other channels for producing new particles are negligible. Processes where the sum of the masses of the final state particles was greater than 5.5 TeV, or 2 TeV for colorless sparticles, were assumed to have negligible cross sections, and the associated event generation was omitted. For each process, the number of events generated was chosen to be equivalent to 25 fb$^{-1}$ of data, subject to a minimum of 100 events and a maximum of 250k events. We impose these lower and upper limits on the number of events for each process to ensure that outliers don't skew our results and to perform event generation within a reasonable amount of time, respectively. In employing PGS, we closely follow the object reconstruction techniques described in the ATLAS MET analyses~\cite{Aad:2011ib, Aad:2011qa, ATLAS:2011ad, Aad:2011cwa, ATLAS:2012aa, ATLAS:2012bb, ATLAS:2012cc}.

At both 7 TeV and 8 TeV, there is a small number of models for which event generation did not complete successfully, usually because Pythia failed to produce the minimum 100 events for a process with a very low cross section within a reasonable amount of time. We omit these models, which make up approximately 0.1\% of the neutralino LSP model set, and do not expect this to materially affect our results.

For each model, we compute K-factors separately for each relevant process involving squarks or gluinos using Prospino 2.1~\cite{Beenakker:1996ch}, at CM energies of both 7 TeV and 8 TeV. Other processes involving only electroweak gauginos, sleptons, and/or Higgs bosons are assigned a K-factor of 1.25.

As Prospino assumes degenerate squark masses for all NLO calculations, and squarks in the pMSSM do not generally have degenerate masses, we note that the K-factors are not calculated using the true squark mass. For the purely strong processes of squark and gluino production, Prospino uses the average of the squark masses as a universal squark mass in the K-factor calculation. These K-factors tend to grow (eventually becoming unstable) with increasing final-state mass, and we thus caution that our NLO cross sections may be somewhat overestimated for squark and gluino production when the various squarks in a model have very different masses. For example, the squark pair production K-factor, calculated assuming degenerate squarks, grows significantly with increasing average squark mass. Since we apply this K-factor to all squarks, the production rate for the lightest squarks is usually overestimated. Since the lightest squarks dominate the total pair production rate, we expect that the total cross section for squark pair production is also overestimated at NLO when the average and lightest squark masses differ significantly.  This will have the effect of increasing the number of
models that are potentially observable at the LHC.

Conversely, for associated squark-neutralino production, we calculate K-factors for the mode where the final state squark is the lightest squark. The resulting K-factors are then used for all possible final state squarks. In this case, our approach is conservative and underestimates the NLO cross section since associated production of a heavier squark with a neutralino would be expected to have a larger K-factor than we employ. Similarly, the K-factors for associated squark-chargino production are calculated using the lightest left-handed squark. We attempted to apply a similarly conservative strategy to the purely strong production processes discussed above, but the ability to use a lighter squark to calculate K-factors for these processes is not currently available in Prospino.

To see the effect of the 2011 LHC data on our model set, we first analyze the 7 TeV events. For each model, we pass the SUSY events from PGS through an analysis code that aims to replicate the ATLAS searches for jets plus missing energy~\cite{Aad:2011ib}, large jet multiplicity~\cite{Aad:2011qa}, one lepton plus jets plus missing energy~\cite{ATLAS:2011ad}, and two leptons plus jets plus missing energy~\cite{Aad:2011cwa}. We scale the events for each process so that they correspond to 1.04 fb$^{-1}$ (0/1/2 leptons + jets + MET) or 1.34 fb$^{-1}$ (large jet multiplicity) of integrated luminosity, corresponding to that used in the ATLAS analyses, and multiply by the appropriate K-factors. We then compare the number of events predicted by our models in each search region of the analyses to the ATLAS 95\% confidence level upper limit on the allowed numbers of signal events in these regions. Models for which the expected number of events in any search region is higher than this 95\% limit are considered to be excluded.

In implementing the ATLAS searches, we have closely followed their published definitions and cuts. For example, we match ATLAS by using the anti-$k_T$ jet algorithm with cone size 0.4, and apply the given $\eta$ and $p_T$ cuts for each object type. However, we have made adjustments in cases where replicating the ATLAS analysis exactly would be very difficult or impossible. For instance, the searches with lepton vetoes define electrons using ``medium'' shower shape and track selection criteria, whereas electrons in search regions with leptons are typically defined with more stringent criteria, including a track momentum isolation requirement. Since we have used a single electron definition in PGS for our event generation, we then later implement an additional isolation requirement in our analysis for the one- and two-lepton search regions based on the momenta of reconstructed objects (not individual tracks). Also, an electronics issue with the ATLAS liquid argon barrel calorimeter was treated in various ways by the MET searches, often using procedures that we are unable to reproduce. Fortunately, this problem is not expected to significantly affect the signal acceptance~\cite{Aad:2011ib}.

\begin{figure}
\centerline{\includegraphics[width=4.5in]{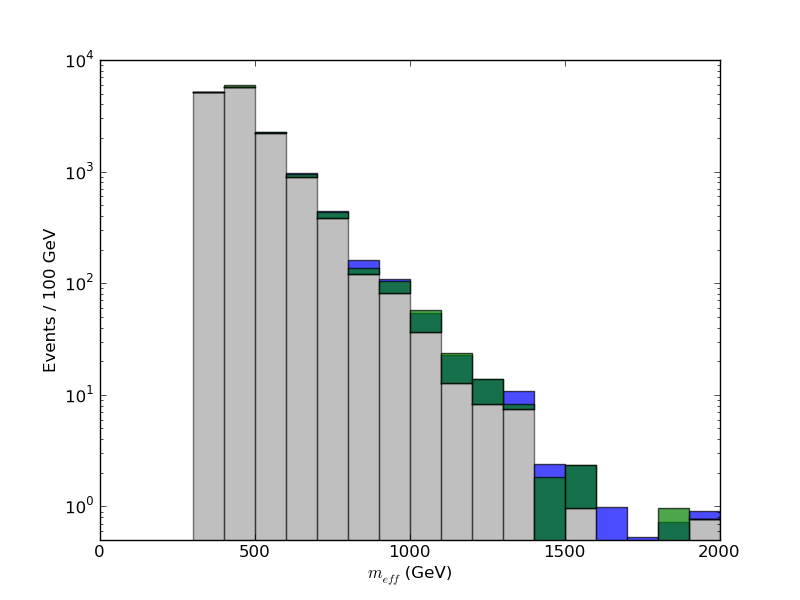}}
\vspace*{0.5cm}
\caption{Effective mass distribution for events passing the cuts of the two jet signal region from the ATLAS jets plus MET search~\cite{Aad:2011ib}. The SM background is shown in gray, with our signal prediction (blue) and the corresponding ATLAS signal prediction (green) on top, for the benchmark mSUGRA point $m_0 = 660\gev, m_\frac{1}{2} = 240\gev, A_0 = 0\gev, \tan \beta = 10, \mu > 0$. Imposing the effective mass cut of 1000 GeV leaves us with 42.2 events, which compares favorably with the ATLAS result of 38.9 events.}
\label{fig:validate}
\end{figure}

Despite slight differences between our analysis routines and those used by the ATLAS collaboration, we achieve reasonable agreement with ATLAS for multiple benchmark models. We have validated our analysis code by generating and analyzing SUSY events for several mSUGRA benchmark points, and then comparing our results with those of ATLAS for these points. In general, our generated signal is in very good agreement with the ATLAS prediction for these benchmark models. As an example, Fig.~\ref{fig:validate} shows SUSY events from the mSUGRA point $m_0 = 660\gev, m_\frac{1}{2} = 240\gev, A_0 = 0\gev, \tan \beta = 10, \mu > 0$, after applying all cuts (except for the effective mass cut) in the two jets plus missing energy search region of the 1.04 fb$^{-1}$ ATLAS analysis~\cite{Aad:2011ib}. We see that our event distribution faithfully reproduces that of ATLAS for this benchmark point.

\begin{table}
\begin{center}
\begin{tabular}{|c||c|c|c|c|}
\hline
Analysis & 7 TeV 1 fb$^{-1}$ & 7 TeV 4.7 fb$^{-1}$ & 8 TeV 5 fb$^{-1}$ & 8 TeV 20 fb$^{-1}$ \\
\hline
\hline
Jets + MET & 6.68\% & 23.23\% & \emph{32.70\%} & \emph{45.11\%} \\
\hline
Many jets + MET & 0.36\% & 1.61\% & \emph{6.26\%} & \emph{7.35\%} \\
\hline
1 $\ell$ + jets + MET & 0.81\% & 2.64\% & \emph{1.41\%} & \emph{1.53\%} \\
\hline
2 $\ell$ + jets + MET & 0.16\% & \emph{0.22\%} & \emph{0.35\%} & \emph{0.38\%} \\
\hline
\hline
Remaining models & 93.27\% & 76.72\% & \emph{67.25\%} & \emph{54.87\%} \\
\hline
\end{tabular}
\caption{The percentages of models excluded (or expected to be excluded, in italics) by each of the ATLAS missing energy analyses, as well as those surviving all of the searches. The ``1 fb$^{-1}$'' 7 TeV searches use 1.34 fb$^{-1}$ of data for the large jet multiplicity search regions and 1.04 fb$^{-1}$ for all other searches. The 8 TeV results are estimates based on the existing 7 TeV searches, as described in the text. The 7 TeV two lepton 4.7 fb$^{-1}$ result is also an estimate based on the 1.04 fb$^{-1}$ two lepton search; it does not significantly change the percentage of models remaining after the other 7 TeV 4.7 fb$^{-1}$ searches.}
\label{tab:excluded}
\end{center}
\end{table}

A summary of the impact of the 7 TeV ATLAS MET-based searches appears in the first column of Table~\ref{tab:excluded}. We find that the jets plus missing energy search with a lepton veto is most effective in excluding models in our set, owing to higher cross sections for production of colored sparticles. Only a small fraction of the model set is excluded by at least one of the search regions. This is generally due to kinematics. Given our choices for the pMSSM parameter scan ranges, it is relatively unlikely for a random model in our scan to have squarks or gluinos light enough to be observed with this amount of data at the LHC. Other potential reasons for models to have been missed by these 7 TeV 1 fb$^{-1}$ ATLAS searches, including compressed sparticle spectra, have been thoroughly discussed~\cite{Strubig:2012qd} in existing studies of the pMSSM.

\begin{figure}
\centering
\subfloat{\includegraphics[height=3.5in]{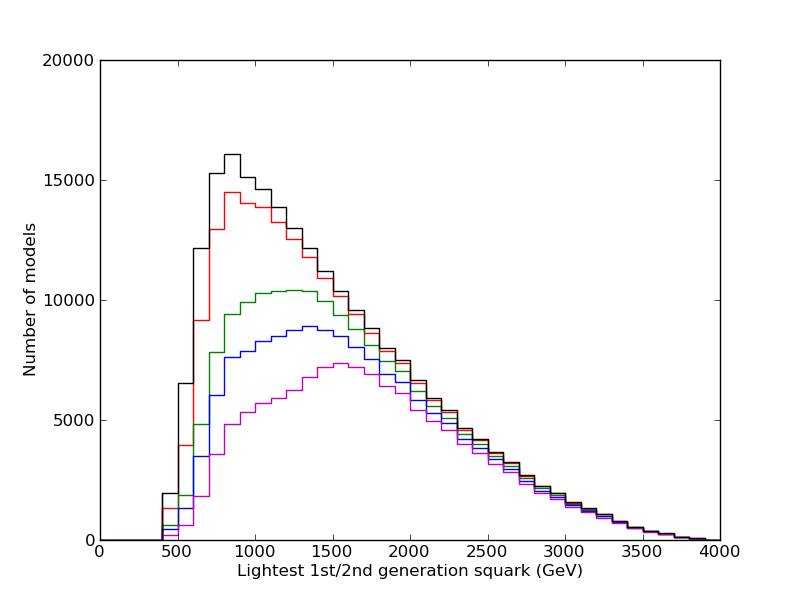}} \\
\subfloat{\includegraphics[height=3.5in]{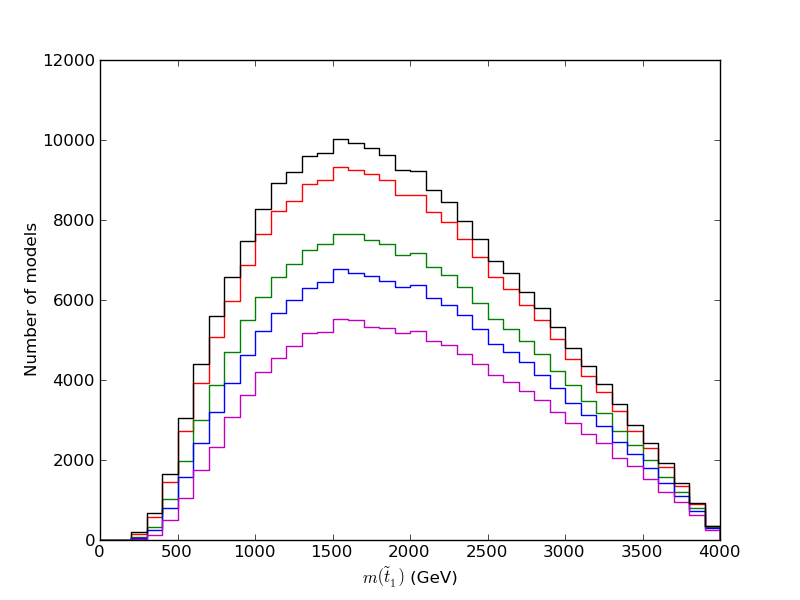}}
\vspace*{0.5cm}
\caption{The distributions of the lightest 1st/2nd generation squark (top) and the lightest stop (bottom) masses among the models in the neutralino LSP set. Results are shown for the original model set (black), as well as for the subsets of models that pass, or are expected to pass, the 7 TeV 1 fb$^{-1}$ (red), 7 TeV 5 fb$^{-1}$ (green), 8 TeV 5 fb$^{-1}$ (blue), and 8 TeV 20 fb$^{-1}$ (magenta) searches.}
\label{fig:excluded1}
\end{figure}

\begin{figure}
\centering
\subfloat{\includegraphics[height=3.5in]{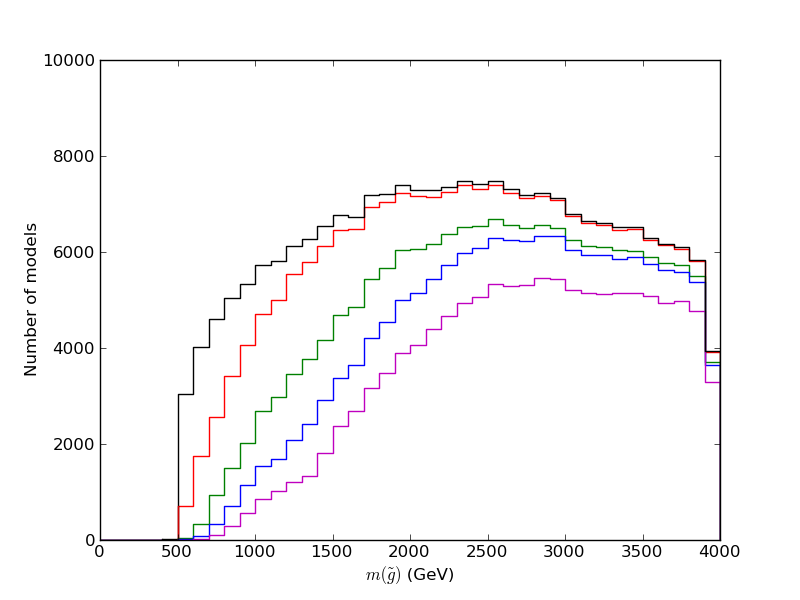}} \\
\subfloat{\includegraphics[height=3.5in]{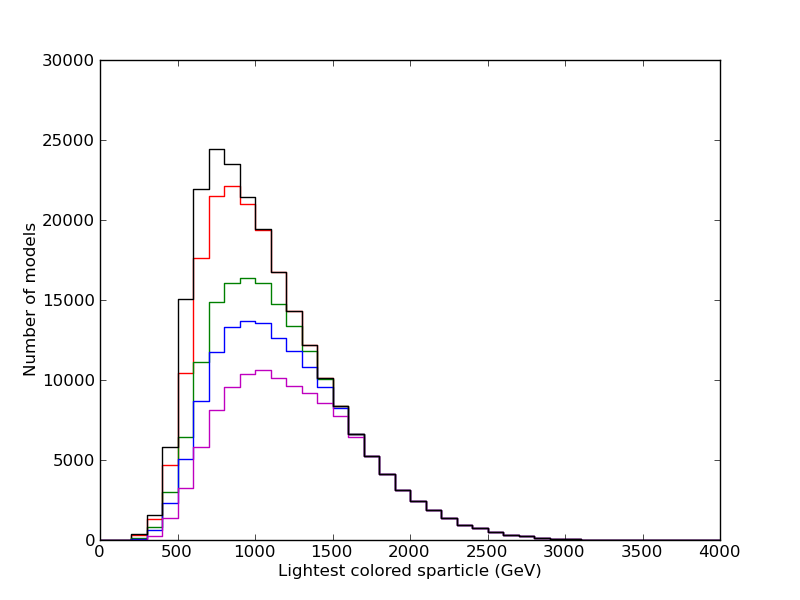}}
\vspace*{0.5cm}
\caption{The distributions of the gluino (top) and the lightest colored sparticle (bottom) masses among the models in the neutralino LSP set. Results are shown for the original model set (black), as well as for the subsets of models that pass, or are expected to pass, the 7 TeV 1 fb$^{-1}$ (red), 7 TeV 5 fb$^{-1}$ (green), 8 TeV 5 fb$^{-1}$ (blue), and 8 TeV 20 fb$^{-1}$ (magenta) searches.}
\label{fig:excluded2}
\end{figure}

\begin{figure}
\centering
\subfloat{\includegraphics[height=3.5in]{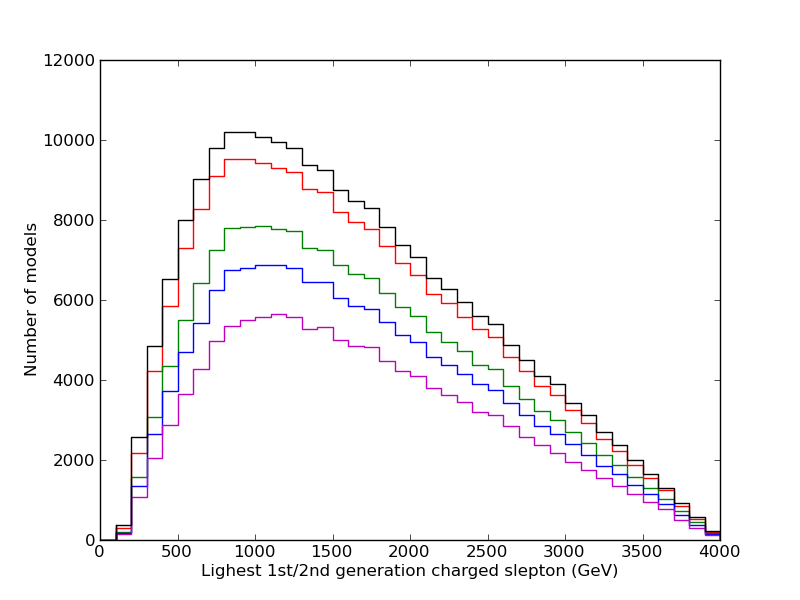}} \\
\subfloat{\includegraphics[height=3.5in]{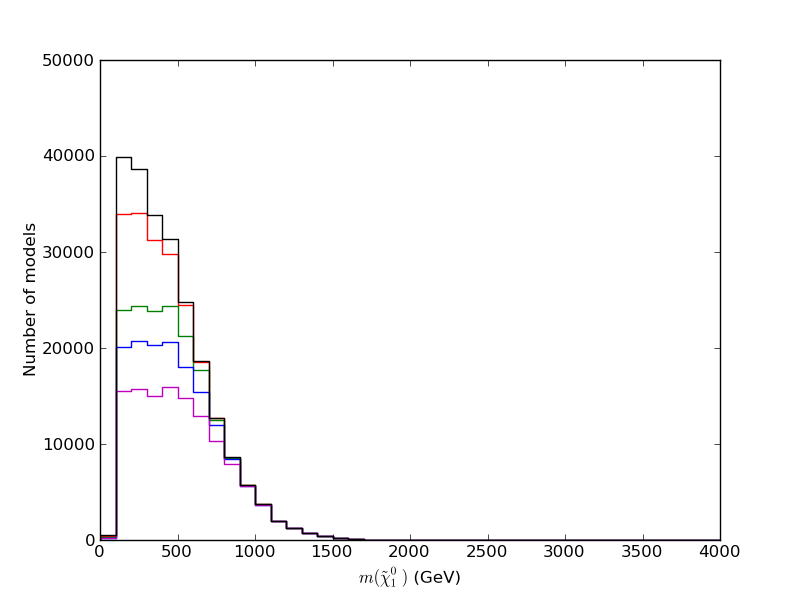}}
\vspace*{0.5cm}
\caption{The distributions of the lightest 1st/2nd generation charged slepton (top) and the lightest neutralino (bottom) masses among the models in the neutralino LSP set. Results are shown for the original model set (black), as well as for the subsets of models that pass, or are expected to pass, the 7 TeV 1 fb$^{-1}$ (red), 7 TeV 5 fb$^{-1}$ (green), 8 TeV 5 fb$^{-1}$ (blue), and 8 TeV 20 fb$^{-1}$ (magenta) searches.}
\label{fig:excluded3}
\end{figure}

Figure~\ref{fig:excluded1} shows the effect of the ATLAS searches on the squark mass distributions of our neutralino model set. Clearly, lighter 1st and 2nd generation squarks are being excluded efficiently, although a few models with light squarks remain.  However, these jets + MET based ATLAS searches have more difficulty observing light stops, as revealed by the similarity of the stop mass distributions before and after the searches we implemented. We expect that other analyses with b-tags will be more useful for excluding third generation squarks~\cite{bjetsearches}. Figure~\ref{fig:excluded2} presents analogous plots for the mass distributions of the gluino and the lightest colored sparticle. These figures indicate that the current ATLAS searches are effective at seeing colored objects and could exclude such particles with mass less than $700-800$ GeV in our model set after the 8 TeV run. On the other hand, it is well-known that the masses of colorless sparticles generally have no bearing on the ability of the LHC to observe a given SUSY spectrum. Figure~\ref{fig:excluded3} shows how the ATLAS searches affect the mass distributions of the charged selectrons/smuons and $\tilde{\chi}_1^0$. The shapes of the slepton mass distribution before and after the 7-8 TeV LHC searches are identical, as expected. Because the LSP mass sets a lower bound on the masses of colored sparticles, however, the ATLAS searches tend to not exclude models with a heavy $\tilde{\chi}_1^0$.

Recently, new ATLAS results that include most of the 2011 data have been presented. These MET searches for supersymmetry look in regions with 0 leptons and either 2-6 jets~\cite{ATLAS:2012aa} or 6-9 jets~\cite{ATLAS:2012bb}, or with 1 lepton~\cite{ATLAS:2012cc}, using 4.7 fb$^{-1}$ of integrated luminosity. The second column of Table~\ref{tab:excluded} shows the results of these searches, as do Figs.~\ref{fig:excluded1}-\ref{fig:excluded3}. These recent analyses tend to have higher missing energy and effective mass cuts, which prove to be very effective in excluding our models that tend to have heavier sparticles. In addition, we include the estimated reach of the 1.04 fb$^{-1}$ 2-lepton ATLAS search~\cite{Aad:2011cwa} if it were to be extended to the 4.7 fb$^{-1}$ data sample with the same search regions. For this extrapolation, we scale the ATLAS backgrounds to account for increased luminosity and then use the CL$_s$ method~\cite{Read:2002hq} to estimate the expected signal limits in each search region, \ie, the limits obtained assuming that the observed number of events matches the expected number of background events in each search. We note that all of the models excluded by the 1.04 fb$^{-1}$ searches are also excluded by the 4.7 fb$^{-1}$ searches. Thus, for sparticles in the mass ranges we have chosen, the tighter cuts of the newer search regions do not introduce gaps in coverage.

Since the LHC collision energy has been increased to 8 TeV proton collisions for 2012, we have also estimated the performance of ATLAS in excluding models in our pMSSM set with both 5 and 20 fb$^{-1}$ of integrated luminosity at this higher energy. To do this, we must estimate the background in certain search regions by extrapolating from the 7 TeV ATLAS data-driven analyses. Specifically, we simulated the Standard-Model background processes $t\bar t +X$, $t+X$, $V+X$ (where $V=W/Z$),
$VV+X$ and pure QCD jets at both 7 TeV and 8 TeV.  All samples were produced with version 1.3.1 of the Sherpa event 
generator~\cite{sherpa}. For the electroweak gauge bosons, as well as 
the top-quarks, we considered both leptonic and hadronic decays. To account for high-$p_T$ 
events and configurations with more than one hard parton inside a large jet, we employ Sherpa's 
matrix-element parton-shower matching algorithm~\cite{Hoeche:2009rj, Hoeche:2009xc}.
We consider tree-level matrix elements with up to four additional final-state partons 
in dijet and $W/Z$-boson production and up to two (one) additional partons in single-top 
($t\bar t$) production.\footnote{Inclusion of the $t\bar t$ + 2 jets background is not expected to alter our results. We have checked this explicitly for the case of the jets + MET analysis, and expect that it will hold for the other analyses as well.} Renormalization and factorization scales are dynamically 
determined on an event-by-event basis according to the matching algorithm~\cite{Hoeche:2009rj}. 
For the parton-separation parameter of the matching procedure we use $Q_{\rm cut} = 20$ GeV,
except in the pure QCD case, where we employ $Q_{\rm cut} = 50$ GeV due to the tight cuts
on multi-jet configurations.

By passing our background events through the cuts for the various searches, we obtain predicted numbers of background events for each search region at both 7 TeV and 8 TeV. Generally, the 7 TeV events effectively reproduce the total numbers of background events observed by ATLAS in these search regions. To extrapolate the 7 TeV backgrounds to 8 TeV, we calculate the ratio of the number of background events at 7 TeV and 8 TeV for each search region, and then use this scaling factor to estimate the 8 TeV background in each region. In taking ratios of our background events, we attempt to limit the effect of systematic errors arising from differences between our background generation procedures and those used by ATLAS. We use these ratios from our monte-carlo simulation to scale the published data-driven backgrounds in the 7 TeV ATLAS analyses to 8 TeV, assuming that all cuts remain unaltered. At 8 TeV, we estimated expected backgrounds with both 5 fb$^{-1}$ and 20 fb$^{-1}$ of integrated luminosity. We then computed the expected limits on the number of signal events using the CL$_s$ method as above under the assumption that the observed background is the same as our calculated background. We find that some limits do not improve significantly, or even become slightly weaker, in cases where the existing 7 TeV searches observed fewer events than the expected background. Models predicting more events in any signal region than our predicted limit are expected to be excluded. For the 2-lepton analysis, we have used the search regions from the 1.04 fb$^{-1}$ ATLAS searches as a baseline, while for all other searches we have extrapolated from the more recent results incorporating 4.7 fb$^{-1}$ of data. 

The 8 TeV results are shown in the last two columns of Table~\ref{tab:excluded} and in Figs.~\ref{fig:excluded1}-\ref{fig:excluded3}. We find that the 8 TeV run has the potential to exclude a significant number of models in addition to those already eliminated by the existing 7 TeV data. For search regions with significant background uncertainty where the calculated scaling factor was large, we note that the expected sensitivity is somewhat poorer than anticipated because of the large errors in the expected 8 TeV backgrounds.  Consequently, if future searches using higher energy and integrated luminosity have lower uncertainty on the measured backgrounds, as may be expected given the use of data-driven techniques for background estimation, then the ability of ATLAS to exclude pMSSM models at 8 TeV will be somewhat greater than our estimates suggest.

\section{LHC Non-MET Searches}
\label{sec:nonmet}

In addition to the SUSY searches involving missing energy that were considered in Section~\ref{sec:met}, several other recent results influence our model set. We consider them here.

\begin{figure}
\centerline{\includegraphics[width=6.5in]{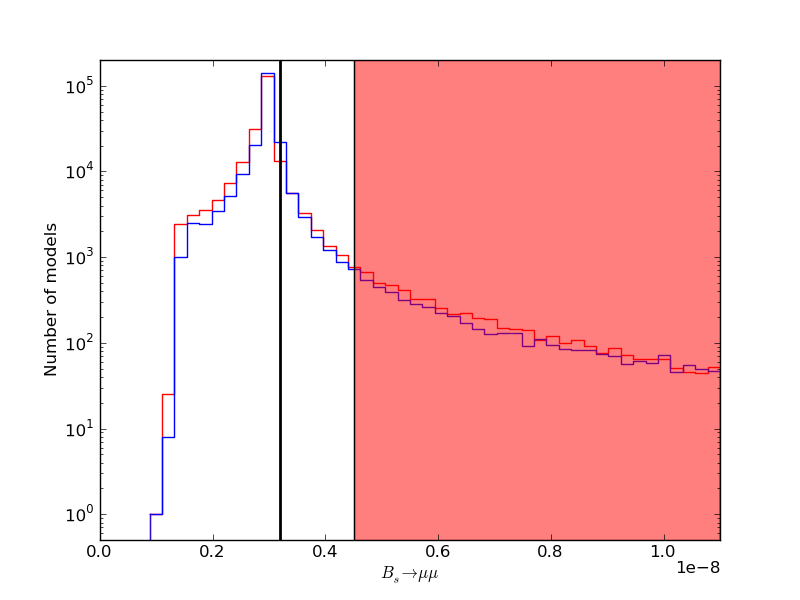}}
\vspace*{0.5cm}
\caption{Values of the $B_s \to \mu \mu$ branching ratio in our neutralino (blue) and gravitino (red) LSP model sets. The black line is the Standard Model prediction, while the shaded region is allowed by the 
present combined LHC limit~\cite{lhcbbs}.}
\label{fig:bsmumu}
\end{figure}

The recent searches for the rare decay $B_s \to \mu \mu$~\cite{atlasbs, cmsbs, lhcbbs} have set a combined limit $\mathrm{BR}(B_s \to \mu \mu) < 4.5 \cdot 10^{-9}$~\cite{lhcbbs}, which
excludes 4899 (5884) of our neutralino (gravitino) LSP models. This represents only $\sim 2\%$ of either model set\footnote{Very recently, ATLAS, CMS AND LHCb have presented a combined upper
limit on the $B_s \to \mu^+\mu^-$ branching fraction of $4.2 \times 10^{-9}$
at the $95\%$ CL{\cite {update}}. This new bound would exclude 6035 (7147)
models within our neutralino (gravitino) model set.}. 
Figure~\ref{fig:spect5} shows the predicted values for $B_s \to \mu^+\mu^-$ for both model sets while Fig.~\ref{fig:bsmumu} shows the 
distributions of $\mathrm{BR}(B_s \to \mu \mu)$ in our model sets with the new limit overlaid. 
As our distributions for flavor observables tend to peak sharply at the SM values, upper bounds on such quantities tend to provide little exclusion power. Eventually, however, precise 
\emph{measurements} of rare decays such as $\mathrm{BR}(B_s \to \mu \mu)$ will be useful in constraining these pMSSM models. It is likely that LHCb will observe this decay mode by the end of 2012. 

\begin{figure}
\centering
\subfloat{\includegraphics[height=3.5in]{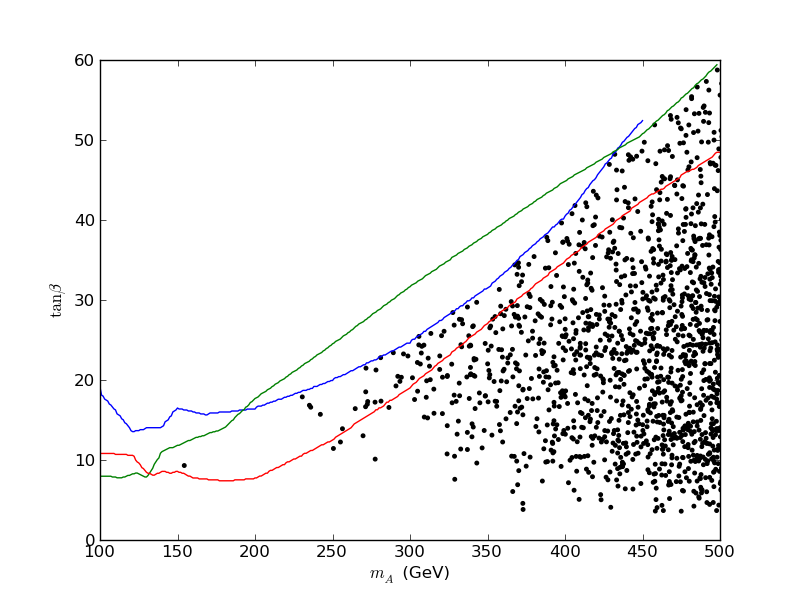}} \\
\subfloat{\includegraphics[height=3.5in]{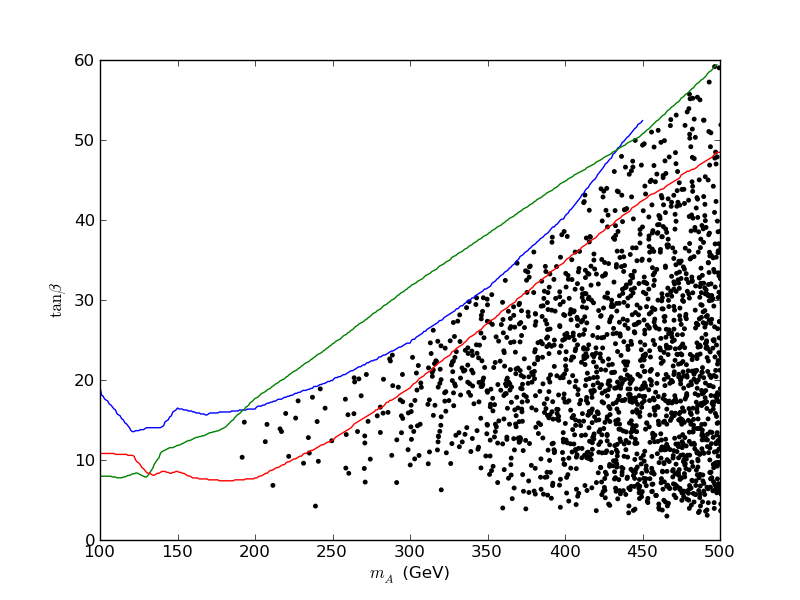}}
\vspace*{0.5cm}
\caption{The effects of $H,A \to \tau \tau$ searches on our neutralino (top) and gravitino (bottom) LSP model sets. We show both the new CMS bound~\cite{Chatrchyan:2012vp} (red), as well as the previous ATLAS (blue) and older CMS 
(green) constraints which were used in generating these model sets.}
\label{fig:matanb}
\end{figure}

The CMS search for neutral Higgs bosons decaying to tau pairs~\cite{Chatrchyan:2012vp} also represents an improvement over the constraints present at the time of model set generation. 
Figure~\ref{fig:matanb} shows this new limit in the $m_A-\tan \beta$ plane, along with the existing limits. Because the new CMS limit only covers a fraction of our $m_A$ range and is only 
slightly better than the previous limits from ATLAS and CMS, its influence on our model sets is minor. 160 (164) models, or 0.07\% of either neutralino (gravitino) set, are excluded by this search.

\begin{figure}
\centering
\subfloat{\includegraphics[height=3.5in]{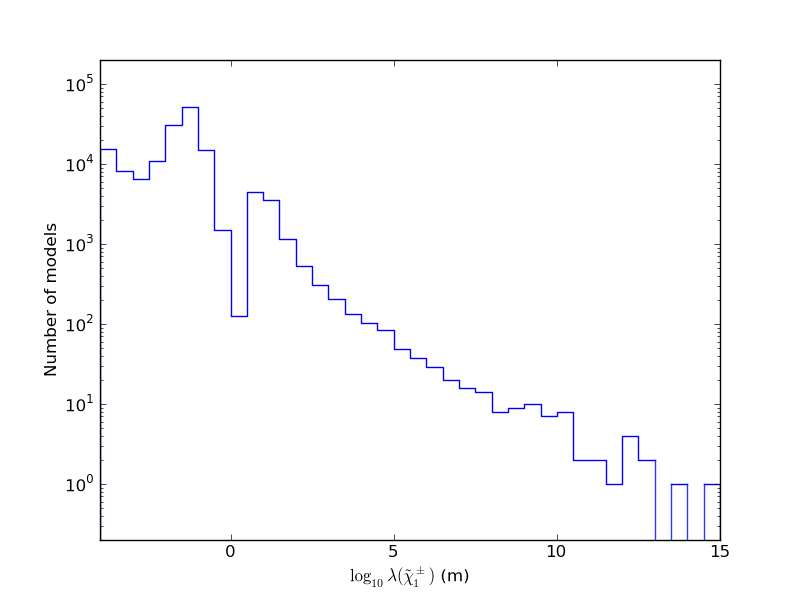}} \\
\subfloat{\includegraphics[height=3.5in]{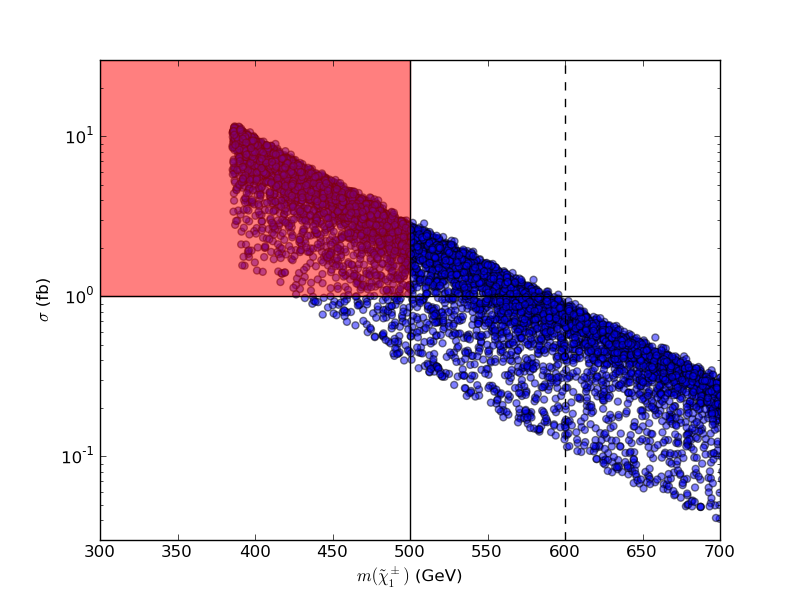}}
\vspace*{0.5cm}
\caption{(Top) Distribution of the log of the unboosted chargino decay length in the neutralino model set. (Bottom) Pair production cross sections for long-lived charginos in the neutralino 
set and the region 
excluded by the CMS stable particle search (red). The dotted line shows the additional exclusion if the CMS bounds were extended to masses of 600 GeV.}
\label{fig:stable1}
\end{figure}

Finally, as we saw in the discussion above, the lightest neutralino mass eigenstate in our model sets tends to be rather close to one of the electroweak eigenstates since the typical 
magnitudes of the parameters $\mu, M_{1,2}$ are of order $\sim 1$ TeV while off-diagonal elements of the neutralino and chargino mass matrices are much smaller. Furthermore, 
as was shown earlier, in the neutralino LSP model set, $\sim 89\%$ of these lightest neutralinos are either wino- or Higgsino-like with high purity implying that the 
corresponding lightest chargino, $\tilde \chi_1^\pm$, is 
likely to be quite nearby in mass. If this mass splitting is below $\sim m_\pi \sim 130$ MeV, the resulting chargino will then be long-lived and will likely appear in the 
detector as a stable charged particle. As described  above, our criterion for a ``detector-stable'' charged particle is that its total decay width satisfy the bound  
$\Gamma < 5 \times 10^{-17}$ GeV. This corresponds to an unboosted decay length of $\sim 4$ m while typically $\gamma\beta\sim 2-3$ are experienced for charginos arising from 
either direct pair production or in decay cascades~\cite{us}. Charginos satisfying this lifetime criterion are relatively abundant in the neutralino LSP model set and occur over a 
wide range of masses above the cut of 385 GeV that was employed in generating these model sets. We find that there are roughly $\sim 10.8$k such neutralino LSP models with their direct pair 
production cross sections at the 7 TeV LHC as shown in Fig.~\ref{fig:stable1}.  

CMS{\cite{CMSstable}} has recently performed a search for the direct production of stable, singly-charged, color-singlet particles with an integrated luminosity of 
$\sim 4.7$ fb$^{-1}$ which they interpret in terms of a long-lived stau. It is possible to re-interpret this limit instead in terms of long-lived charginos
{\cite{Brooijmans:2012yi}}. For charginos in the $\sim 385-500$ GeV mass range this limit corresponds to an upper bound of $\simeq 1$ fb on the direct chargino pair production 
cross section. Examining the lower panel in Fig.~\ref{fig:stable1}, we see that there are many neutralino pMSSM models in this mass range ($\sim3.6$k) that are excluded by this new stable 
particle search. We further observe that if the CMS limit were to be extended to a mass of $\sim 600$ GeV, $\sim 1.4$k additional models would also be excluded. 

This search provides further evidence for the power of non-MET searches to explore the pMSSM model sets.

\section{Conclusion}
\label{sec:conc}

The pMSSM allows us to investigate complete, realistic Supersymmetric spectra at the LHC and beyond.  In this paper, we have updated and extended our earlier pMSSM analyses by
generating two new, very large model samples, one set with a neutralino LSP and the second containing a gravitino LSP.  These model sets have been subjected to a set of global
constraints arising from theoretical considerations, cosmological observations, and data from the precision electroweak sector, the flavor sector as well as LEP/Tevatron searches.
Approximately 225,000 models were found to be viable for each set, thus enlarging our exploration of the pMSSM parameter space.
In these new model samples, the mass range of the sparticles has
been extended to an upper limit of 4 TeV, to allow for a robust examination of the pMSSM properties as the LHC data comes in and so that a good fraction of the model sample will remain 
viable for the 14 TeV LHC.  These two model sets can then be used as the basis for numerous investigations for the years to come.  

The consideration of a gravitino LSP necessitated a careful examination of the gravitino's cosmological effects.  In this case, the potentially long-lived NLSP plays a
meaningful role in the cosmology.  The computation of the LSP relic abundance differs substantially from the scenario where the LSP is the lightest neutralino and, here, we considered non-thermal production based solely upon the relic density of the NLSP.  We found that the WMAP bounds placed the strongest constraints
on models where the NLSP was a sneutrino.  Big Bang Nucleosynthesis was found to apply sharp restrictions on the gravitino model set for NLSP lifetimes $\gsim 10^{2-3}$ s, 
particularly in the case where the NLSP resulted in a charged or hadronic energy injection.  The BBN bounds on sneutrino NLSPs were found to be weaker and could be enhanced by
an improvement in the evaluation of BBN.  The diffuse photon and neutrino spectra were found to have little effect in constraining our model set.

We next compared the characteristics of the two models sets and found contrasts in some cases.  In the neutralino LSP set, the lightest chargino (and the second lightest neutralino)
are by far the most frequent NLSP, whereas in the gravitino LSP set, the identity of the NLSP is essentially arbitrary.  However, the influence of stable charged particle searches
and BBN result in a slight preference for non-colored states to be the NLSP in the gravitino case.  The lightest neutralino and chargino masses favor towards smaller values in the
neutralino LSP set (as the $\tilde\chi_1^\pm$ tends to track the $\tilde\chi_1^0$), while the mass distribution for these sparticles in the gravitino models is essentially flat.  
Stop masses tend to be somewhat lower in the gravitino LSP set, resulting in a shift to smaller values for the lightest CP-even Higgs mass.  The implications of this result will
be studied further \cite{future}. Predictions for processes such as $b\to s\gamma,\ B_s\to\mu^+\mu^-$ and the anomalous magnetic moment of the muon
were found to be similar between the gravitino and neutralino sets.

The effects of the 1 and 5 fb$^{-1}$ 7 TeV ATLAS MET-based SUSY searches were examined in detail for the neutralino LSP model sample and found to already exclude roughly 23\% of the 
models in our set. Here, we closely followed the published ATLAS analyses. The availability of data-driven background measurements (versus MC estimates), as well as a 
large set of benchmark models validated by ATLAS, served to sharpen the accuracy of our predictions.
We also extrapolated these results to estimate the pMSSM coverage at the current 8 TeV run of the LHC.  
We performed this extrapolation by computing a MC estimate (based on Sherpa) on the ratio of backgrounds from 7 TeV to 8 TeV and then rescaled the measured 7 TeV ATLAS backgrounds.
With this procedure, we found that $5-20$ fb$^{-1}$ at 8 TeV can probe $32-45\%$ of our neutralino LSP model set, resulting in a sizeable extension of coverage over 7 TeV.  
If the pMSSM is not discovered during the 8 TeV run, then our model
set will be essentially void of gluinos and lightest first and second generation squarks that are $\lsim 700-800$ GeV, which is much less than the analogous mSUGRA bound.  
We plan to implement the targeted ATLAS third generation squark searches on our model set in a future study \cite{future}.

Lastly, we saw that non-MET based LHC searches play a significant role in exploring the pMSSM parameter space.  The stable charged particle search is particularly important in the case of the
gravitino LSP model set due to the plethora of long-lived NLSPs.  The details of searching for the gravitino LSP set at the LHC differ from those of the neutralino case, and warrants
an independent comprehensive study; this will be forthcoming \cite{future}.

The pMSSM provides a solid base from which to perform a variety of SUSY studies and to compare the sensitivity of various search techniques.  Here, we have generated two new and
very large pMSSM databases which will have lasting value.  We would prefer SUSY to be discovered at the LHC this year, but if not, these two model sets can be used for years to come. 
 
\section*{Acknowledgments}

The authors are grateful to John Conway for providing an updated version of PGS, and for discussions with J.~Conley, R.~Cotta, T.~Eifert, J.~Gainer, M.~P.~Le, and T.~Plehn. SH's work was 
supported in part by the National Science Foundation, grant NSF-PHY-0705682 (The LHC Theory Initiative).

\newpage

%
\def\IJMP #1 #2 #3 {Int. J. Mod. Phys. A {\bf#1},\ #2 (#3)}
\def\MPL #1 #2 #3 {Mod. Phys. Lett. A {\bf#1},\ #2 (#3)}
\def\NPB #1 #2 #3 {Nucl. Phys. {\bf#1},\ #2 (#3)}
\def\PLBold #1 #2 #3 {Phys. Lett. {\bf#1},\ #2 (#3)}
\def\PLB #1 #2 #3 {Phys. Lett. B {\bf#1},\ #2 (#3)}
\def\PR #1 #2 #3 {Phys. Rep. {\bf#1},\ #2 (#3)}
\def\PRD #1 #2 #3 {Phys. Rev. D {\bf#1},\ #2 (#3)}
\def\PRL #1 #2 #3 {Phys. Rev. Lett. {\bf#1},\ #2 (#3)}
\def\PTT #1 #2 #3 {Prog. Theor. Phys. {\bf#1},\ #2 (#3)}
\def\RMP #1 #2 #3 {Rev. Mod. Phys. {\bf#1},\ #2 (#3)}
\def\ZPC #1 #2 #3 {Z. Phys. C {\bf#1},\ #2 (#3)}

\end{document}